\documentclass[10pt,conference]{IEEEtran}
\IEEEoverridecommandlockouts
\usepackage{cite}

\usepackage{graphicx}
\usepackage{textcomp}
\usepackage{xcolor}
\usepackage{graphicx}
\usepackage{subfigure}
\usepackage{colortbl}
\usepackage{pifont}
\usepackage{multirow}
\usepackage{enumitem}
\usepackage{amsmath, amsfonts}
\usepackage{listings}
\usepackage{xcolor}
\usepackage{framed}
\usepackage{fancyhdr}
\usepackage{url}
\usepackage{pifont}
\usepackage[noend]{algpseudocode}
\usepackage{algorithm}
\usepackage{xspace}
\usepackage{booktabs}

\usepackage{wrapfig}
\usepackage{hyperref}
\hypersetup{
colorlinks = true, 
linkcolor=black, 
citecolor=black, 
urlcolor=black, 
filecolor=black
}

\def\BibTeX{{\rm B\kern-.05em{\sc i\kern-.025em b}\kern-.08em
    T\kern-.1667em\lower.7ex\hbox{E}\kern-.125emX}}

\newcommand\Algphase[1]{%
\vspace*{-.7\baselineskip}\Statex\hspace*{\dimexpr-\algorithmicindent-2pt\relax}\rule{.45\textwidth}{0.4pt}%
\Statex\hspace*{-\algorithmicindent}\textbf{#1}%
\vspace*{-.7\baselineskip}\Statex\hspace*{\dimexpr-\algorithmicindent-2pt\relax}\rule{.45\textwidth}{0.4pt}%
}

\usepackage[switch]{lineno}
\usepackage{todonotes}

\newcommand{\mixcode}{\textsc{MixCode}\xspace}
    
\begin{document}

\title{\mixcode: Enhancing Code Classification by Mixup-Based Data Augmentation}
\author{\IEEEauthorblockN{Zeming Dong\IEEEauthorrefmark{2},
Qiang Hu\thanks{*Qiang Hu is the corresponding author.}\IEEEauthorrefmark{3}\IEEEauthorrefmark{1}, 
Yuejun Guo\IEEEauthorrefmark{4}, 
Maxime Cordy\IEEEauthorrefmark{3}, \\
Mike Papadakis\IEEEauthorrefmark{3},
Zhenya Zhang\IEEEauthorrefmark{2},
Yves Le Traon\IEEEauthorrefmark{3}, and
Jianjun Zhao\IEEEauthorrefmark{2}}
\IEEEauthorblockA{\IEEEauthorrefmark{2}Kyushu University, Japan, \href{mailto:dong.zeming.011@s.kyushu-u.ac.jp}{dong.zeming.011@s.kyushu-u.ac.jp}, \href{mailto:zhang@ait.kyushu-u.ac.jp, zhao@ait.kyushu-u.ac.jp}{\{zhang, zhao\}@ait.kyushu-u.ac.jp}\\
\IEEEauthorrefmark{3}University of Luxembourg, Luxembourg, \href{mailto:qiang.hu@uni.lu, maxime.cordy@uni.lu, michail.papadakis@uni.lu, Yves.LeTraon@uni.lu}{\{qiang.hu, maxime.cordy, michail.papadakis, Yves.LeTraon\}@uni.lu}\\
\IEEEauthorrefmark{4}Luxembourg Institute of Science and Technology, Luxembourg, \href{mailto:yuejun.guo@list.lu}{yuejun.guo@list.lu}}
}

\maketitle
\thispagestyle{fancy}
\pagestyle{fancy}
\cfoot{\thepage}
\renewcommand{\headrulewidth}{0pt} 
\renewcommand{\footrulewidth}{0pt} 

\begin{abstract}


Inspired by the great success of Deep Neural Networks (DNNs) in \emph{natural language processing (NLP)}, DNNs have been increasingly applied in \emph{source code analysis} and attracted significant attention from the software engineering community. Due to its data-driven nature, a DNN model requires massive and high-quality labeled training data to achieve expert-level performance. Collecting such data is often not hard, but the labeling process is notoriously laborious. 
The task of DNN-based code analysis even worsens the situation because source code labeling also demands sophisticated expertise. Data augmentation has been a popular approach to supplement training data  in domains such as computer vision and NLP. However, existing data augmentation approaches in code analysis adopt simple methods, such as data transformation and adversarial example generation, thus bringing limited performance superiority.  In this paper,  we propose a data augmentation approach \mixcode that aims to effectively supplement valid training data, inspired by the recent advance named \emph{Mixup} in computer vision. Specifically, we first utilize multiple code refactoring methods to generate transformed code that holds consistent labels with the original data. Then, we adapt the Mixup technique to mix the original code 
with the transformed code to augment the training data. We evaluate \mixcode on two programming languages (Java and Python), two code tasks (problem classification and bug detection), four benchmark datasets (\emph{JAVA250}, \emph{Python800}, \emph{CodRep1}, and \emph{Refactory}), and seven model architectures (including two pre-trained models \emph{CodeBERT} and \emph{GraphCodeBERT}). Experimental results demonstrate that \mixcode outperforms the baseline data augmentation approach by up to 6.24\% in accuracy and 26.06\% in robustness. 
\end{abstract}
  
\maketitle

\section{Introduction}
\label{sec:intro}
Due to its remarkable performance, deep learning (DL) has gained widespread adoption in different application domains, such as face recognition~\cite{wang2021deep}, language translation~\cite{dong2015multi}, video games~\cite{vinyals2019grandmaster}, and autonomous driving~\cite{vinyals2019grandmaster}. More recently, researchers from the software engineering community have attempted to use DL techniques to automate multiple downstream code tasks, e.g., code search~\cite{gu2018deep}, problem classification~\cite{puri2021codenet}, and bug detection~\cite{shi2021more}. Relevant studies~\cite{feng2020codebert, ma2021graphcode2vec} reveal that DL benefits source code analysis.

As the key pillar of DL systems, deep neural networks (DNNs) automatically gain knowledge from training data and make inferences for unseen data after deployment. Generally, two important factors could affect the performance of the trained DNNs, namely, the \emph{model architecture} and the \emph{training data.} 
In the context of code analysis, for the former factor, a common practice of building proper model architectures of DNNs is to directly apply natural language processing (NLP) models to source code. For example, Feng et al.~\cite{feng2020codebert} have modified BERT to create \emph{CodeBERT} that solves downstream tasks effectively. For the latter factor, though adequate labeled training data are necessary for the training process, producing high-quality labeled source code data is not yet sufficiently investigated. The main challenge is that data labeling requires not only extensive human efforts but also sophisticated domain knowledge. According to to~\cite{zhou2019devign}, labeling code from only four libraries can take up to 600 man-hours. In a nutshell, data preparation is indispensable but challenging for developing desirable models, and therefore in this paper, we take a specific focus on this important issue.

Data augmentation is a technique that tackles the aforementioned data labeling issue, which produces additional training data by modifying existing data rather than human efforts. Generally, the new data sample is semantically consistent with the source data, i.e., they share the same functionalities and labels. In this way, the model can learn more information and gain better generalization, compared to the approach that relies only on the original training data. In computer vision and NLP tasks, data augmentation has been widely used and well-studied~\cite{shorten2019survey,feng2021survey,zhao2021data} for model training. For example, in computer vision tasks, many image transformation methods (e.g., image rotation, shear) are designed to mimic different real-world situations that the model could face after deployment. In traditional NLP tasks, the typical augmentation is to perform synonym substitution, which is also beneficial to cover more context that might occur in the real world. 

Although data augmentation has proved to be effective in fields such as CV and NLP, the investigation of its application in code analysis still remains at an early stage. Researchers have borrowed ideas from other fields and proposed several data augmentation approaches for code analysis~\cite{allamanis2021self,pour2021search, yefet2020adversarial,bui2021self,wang2022bridging}; usually, these techniques generate more transformed or adversarial data simply via methods such as code refactoring. However, existing studies~\cite{yu2022data} already show that these simple strategies have limited effects. For example, Bielik et al.~\cite{pmlr-v119-bielik20a} show that adversarial training, by simply adding adversarial data in the training set, is not helpful in improving the generalization property of DNN models. Therefore, it still remains an open problem to design data augmentation approaches that can effectively enhance DNN training for code analysis. 

In this paper, for source code classification tasks, we propose a novel data augmentation framework named \mixcode that aims to enhance the DNN model training process. Roughly speaking, \mixcode follows a similar paradigm to \emph{Mixup}~\cite{zhang2017mixup} but adapts the technique in order to handle the specific data type of source code. Mixup is a well-known data augmentation technique originally proposed for image classification, which linearly mixes training data, including their labels, to increase the data volume. In our case, \mixcode consists of two steps: first, it generates transformed data by different code refactoring methods, and then, it linearly mixes the original code and transformed code to generate new data.
Specifically, we study 18 types of existing code refactoring methods, such as argument renaming and statement enhancement. More details are in Section~\ref{sec:refactorMethods}.

We conduct experiments to evaluate the effectiveness of \mixcode on two programming languages (Java and Python), two widely-studied code learning tasks (problem classification and bug detection), and seven model architectures. Based on that, we answer three research questions as follows:

\smallskip
\noindent\textbf{RQ1: How effective is \mixcode for enhancing the accuracy and robustness of DNNs?} We compare \mixcode to the standard training (without data augmentation) and the existing simple code data augmentation approach, which relies on transformed or adversarial data only. The results show that \mixcode outperforms the baselines by up to 6.24\% in accuracy and 26.06\% in robustness. Here, accuracy is the basic metric that measures the effectiveness of the trained DNN models.
Moreover, robustness~\cite{xu2012robustness} reflects the generalization ability of the trained model to handle unseen data, which is also an important metric for the deployment of DNN models in practice~\cite{kawaguchi2017generalization}. 

\smallskip
\noindent\textbf{RQ2: How do different Mixup strategies affect the effectiveness of \mixcode?} We study the effectiveness of \mixcode under different settings of Mixup. First, we study the effect of using different data mixture strategies, which involve 1) mixing only original code, 2) mixing original code and transformed code, and 3) mixing only transformed code. Moreover, we also study the effect of different hyperparameters in \mixcode. Our evaluation demonstrates that using the 2) strategy, namely, mixing original code and transformed data, in Mixup can achieve the best performance, and we also make the suggestion on the use of the most suitable hyperparameters of \mixcode. 

\smallskip
\noindent\textbf{RQ3: How does the refactoring method affect the effectiveness of \mixcode?} To investigate the impact of the code refactoring methods on \mixcode, we evaluate \mixcode using different refactoring methods individually. We find that there is a trade-off between the original test accuracy and robustness when choosing different refactoring methods, i.e., using the refactoring methods that lead to higher accuracy could harm the model's robustness.

In summary, the main contributions of this paper are:
\begin{itemize}
    \item We propose \mixcode, the first Mixup-based data augmentation framework for source code analysis. Experimental
results demonstrate that \mixcode outperforms the baseline
data augmentation approach by up to 6.24\% in accuracy and
26.06\% in robustness. The implementation of \mixcode are available online.\footnote{\url{https://github.com/zemingd/Mixup4Code}\label{site}}

    \item We empirically demonstrate that simply mixing the original code is not the best strategy in \mixcode. In addition, \mixcode using original code and transformed code can achieve 9.23\% performance superiority in accuracy.
    
    \item We empirically show that selection of refactoring methods is also an important factor affecting the performance of \mixcode. 
    
\end{itemize}


\section{Preliminaries}
\label{sec:background}

We briefly introduce the preliminaries  of this work from the perspectives of DNNs for source code analysis, DNN model training methods, and Mixup for data augmentation.

\subsection{DNNs for Source Code Analysis}
DNNs have been widely used in NLP and achieved great success. Similar to the natural language text, source code also consists of discrete symbols that can be processed as sequential or structural data fed into DNN models. Thus, researchers have tried to employ DNNs to help programmers process and understand source code in recent years. The impressive performance of DNNs has been demonstrated in multiple important code-related tasks, such as automated program repair~\cite{li2020dlfix, gupta2017deepfix, bhatia2018neuro, pu2016sk_p, chen2019sequencer,yasunaga2020graph, dinella2020hoppity,chen2021plur}, automated program synthesis~\cite{hendrycks2021measuring,zhang2018neural}, and automated code comments generation~\cite{hu2018deep}. 

To unlock the potential of DNNs for code-related tasks, properly representing snippets of code that are fed into DNNs is necessary. Code representation, which transfers the raw code to machine-readable data, plays an important role in source code analysis. Existing representation techniques can be roughly divided into two categories, namely, \emph{sequence representation}~\cite{alon2019code2vec} and \emph{graph representation}~\cite{ma2021graphcode2vec, guo2020graphcodebert,allamanis2017learning}. Sequence representation converts the code into a sequence of tokens. The input features of classical neural networks in sequence representation learning are typically embedded or features that live in Euclidean space. In this way, the original source code is processed to multiple tokens, e.g., from ``\texttt{def func(a, b)}'' to ``\texttt{[def, func, (, a, b, ),]}''. Sequence representation is useful for learning the semantic information of the source code because it remains the context of the source code. On the other side, graph representation builds structural data. In source code, the structure information can be presented by the \emph{abstract syntax tree (AST)} and different code flows (control follow, data flow). By learning these structural data, the model can perceive functional information of code. Recently, more researchers have focused on the field that applies graph representation to source code analysis based on different variants of \emph{graph neural networks (GNNs)}. In our study, we consider both categories of code representation to evaluate \mixcode.

\subsection{DNN Model Training Methods}
DNN training consists in, given a set of training data, searching for the best parameters (e.g., weights, biases) that enable the model to fit the training data. Here, we introduce the standard training process and the basic data augmentation framework for the source code model. Algorithm~\ref{alg:basic} presents the pseudocode of these two training strategies.
In the standard manner of training, all the training data are fed into several epochs of training (See Lines~\ref{line:standardStart}-\ref{line:standardEnd} in Algorithm~\ref{alg:basic}). 

\begin{algorithm}
\caption{Existing model training strategies}\label{alg:basic}
\begin{algorithmic}[1]
\Require $M$: initialized DNN model
\Require $X, Y$: original training data and labels
\Require $R=\left\{r\right\}:$ a set of data transformation methods
\Ensure $M$: trained model
\Algphase{\textcolor{brown}{Standard training (without augmentation)}}
\For{$run \in \{0,\ldots, \#epochs\}$}\label{line:standardStart}
\State $M.\mathsf{Fit}\left(X, Y\right)$
\EndFor
\State \Return $M$\label{line:standardEnd}
\Algphase{\textcolor{brown}{Basic augmentation}}
\For{$run\in \{0, \ldots, \#epochs\}$}\label{line:basicAugStart}
\State $X_{ref}\gets\phi$
\For{$x\in X$}
\State $r\gets \mathsf{RandomSelection}\left(R\right)$
\State $X_{ref}\gets X_{ref}\cup r\left(x\right)$
\EndFor
\State $M.\mathsf{Fit}\left(X_{ref}, Y\right)$
\EndFor
\State \Return $M$\label{line:basicAugEnd}
\end{algorithmic}
\end{algorithm}

However, since the prepared training data can only represent a limited part of data distribution, the training data volume has been a bottleneck that prevents DNN models from achieving high performance~\cite{koh2021wilds}. Data augmentation is proposed to automatically increase the volume of the training set, and thus enhance the quality of training.  The basic idea of data augmentation is to generate new data from the existing training data by well-designed data transformation methods. Generally, such data transformation methods modify the data and do not change their semantic information; for example, in image data processing, commonly-used methods include random rotating, padding, and adding brightness~\cite{shorten2019survey}. 
Line~\ref{line:basicAugStart}-\ref{line:basicAugEnd} in Algorithm~\ref{alg:basic} shows the process of training with data augmentation.
Specifically, in each epoch, the DNN is trained by using a transformed version of the data generated by randomly selected data transformation methods.


\subsection{Mixup: A Data Augmentation Approach in Image Classification Tasks}\label{sec:originMixup}
Mixup~\cite{zhang2017mixup} is an effective data augmentation technique proposed for image classification tasks. Mixup contains two steps: first, it randomly selects two data samples from the training data; then, it mixes both the data features and the labels of the selected data to generate a new sample as the training data. In addition to image classification, recently, researchers have achieved great success in applying Mixup to text classification~\cite{zhu2019mixup}. 

\begin{figure}[!ht]
    \centering
    \subfigure[original image 1]{
    \includegraphics[scale=0.13]{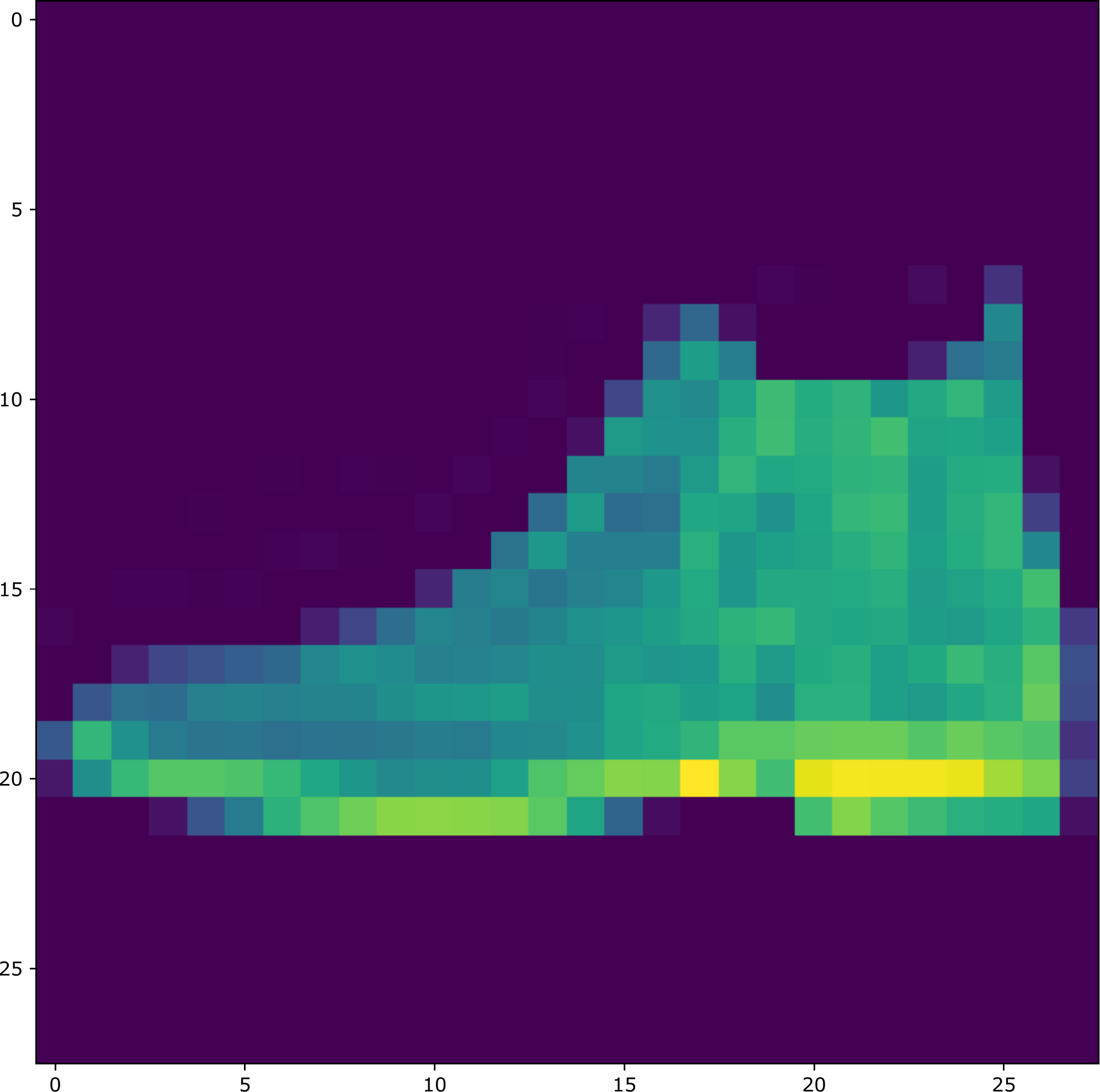}%
    }
    \subfigure[original image 2]{
    \includegraphics[scale=0.13]{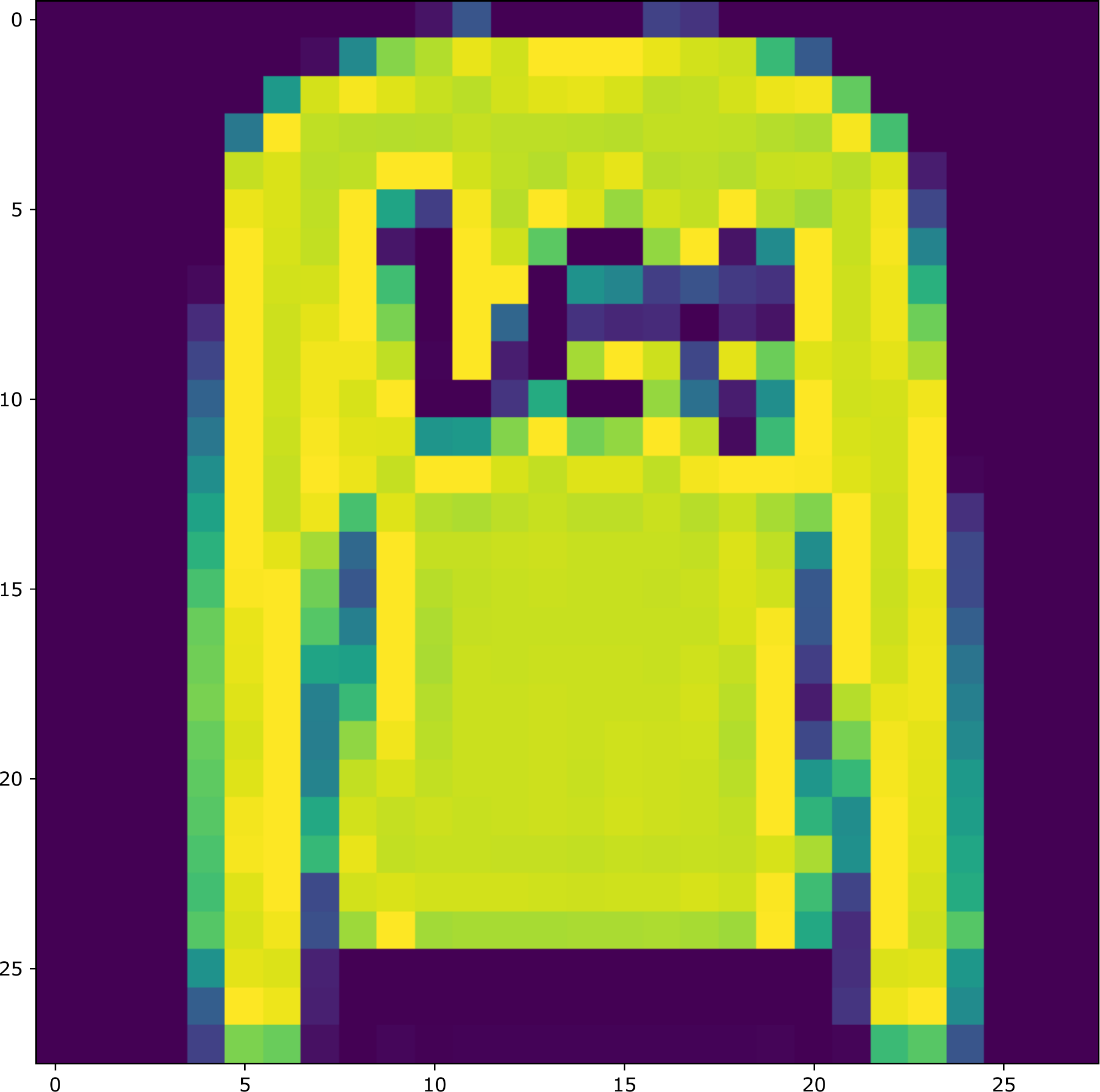}%
    }
    \subfigure[mixed image]{
    \includegraphics[scale=0.13]{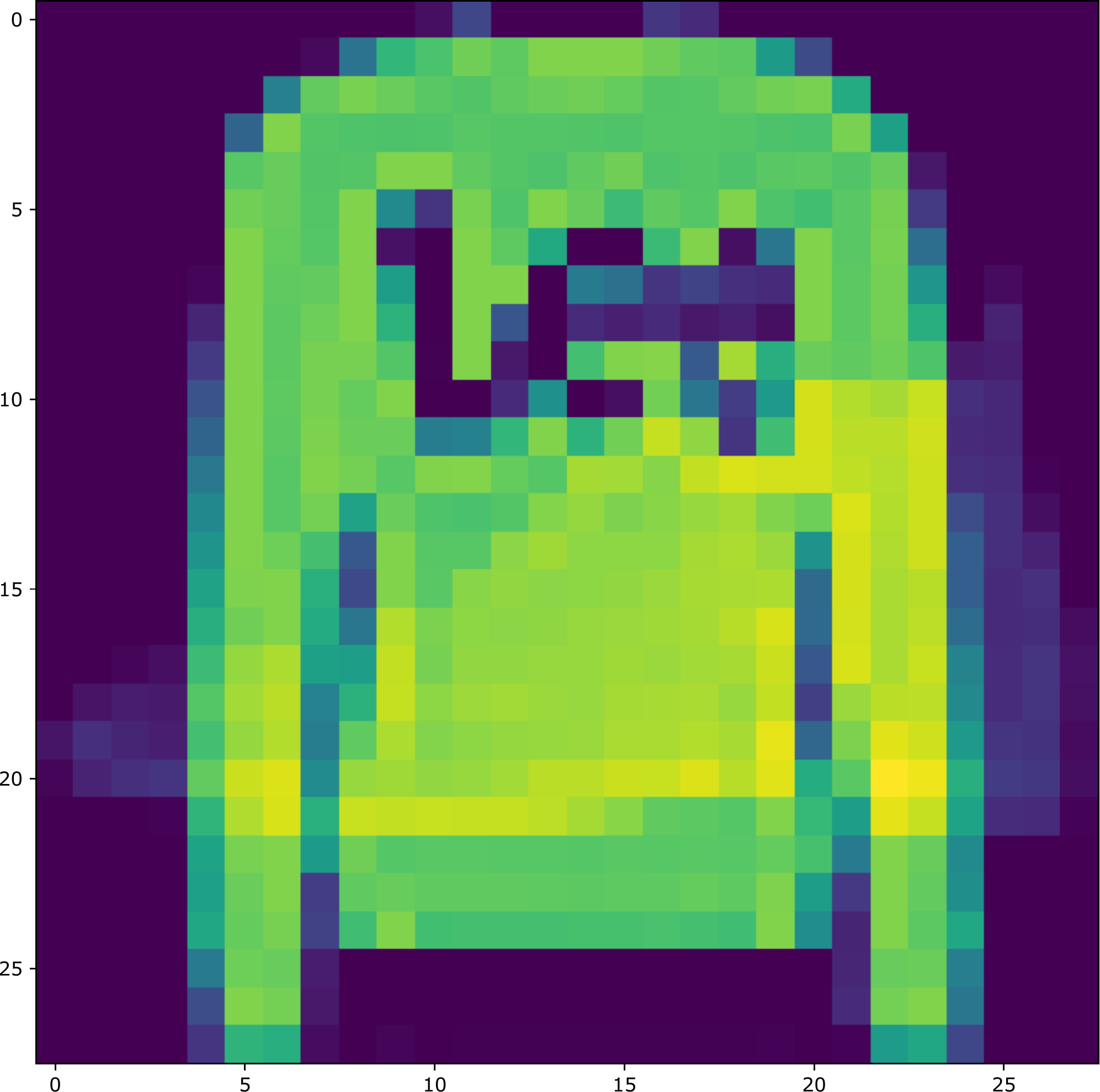}%
    }
    \caption{An example of Mixup for image data. The mixed image is calculated using Eq.~(\ref{eqn:mixup}). $\lambda=0.2$.}
    \label{fig:mix_example}
\end{figure}

Technically, Mixup is shown as follows: given a pair of samples ($x^{i}$,$y^{i}$) and ($x^{j}$,$y^{j}$), where $x$ represents the input feature and its corresponding output label $y$ is donated with one-hot encoding, Mixup produces new data pairs ($x_{mix}^{ij}, y_{mix}^{ij}$):

\begin{equation}\label{eqn:mixup}
\begin{aligned}
    x_{mix}^{ij} = \lambda x^{i} + (1 - \lambda) x^{j}\\
    y_{mix}^{ij} = \lambda y^{i} + (1 - \lambda) y^{j}
\end{aligned}
\end{equation}
where $\lambda$ is a mixing policy for the input sample pair, which is sampled from a \emph{Beta} distribution with a shape parameter $\alpha$ ($\lambda$ $\sim$ \textit{Beta}($\alpha$,$\alpha$)). Figure~\ref{fig:mix_example} depicts an example of an  image generated by Mixup. By mixing two images into one, a model can gain knowledge from both sides.

\section{MixCode---The Proposed Approach}
\label{sec: mixcode}


\subsection{Methodology of \mixcode}

 Inspired by the great success of Mixup and its variants in image classification tasks, we propose \mixcode, a simple yet effective data augmentation framework for source code classification tasks. Algorithm~\ref{alg:mixcode} presents the whole process of \mixcode.
\begin{algorithm}
\caption{\mixcode}\label{alg:mixcode}
\begin{algorithmic}[1]
\Require $M$: initialized DNN model
\Require $T$: code representation technique
\Require $X, Y$: original training data and labels
\Require $R=\left\{r\right\}:$ a set of refactoring methods
\Require $\alpha:$ Mixup weight
\Ensure $M$: trained model
\For{$run\gets \{0, \ldots, \#epochs\}$}
\State $X_{{ref}}, X_{{mix}}, Y_{mix} \gets \phi, \phi, \phi$ \Comment{initialization} \label{line:mixInit}
\For{$x\in X$} \label{line:mixRefactStart}
\State $r\gets \mathsf{RandomSelection}\left(R\right)$ \label{line:mixRandomSelection}
\State $X_{ref}\gets X_{ref}\cup r\left(x\right)$ \Comment{code refactoring}
\label{line:mixRefactAdd}
\EndFor
\State $X_s, Y_s\gets \mathsf{Shuffle}\left(X, Y\right)$ \Comment{Shuffle the training set }\label{line:mixShuffle}
\For{$\left(x_s, y_s, x_{ref}, y\right) \in \left(X_s, Y_s, X_{ref}, Y\right)$}\label{line:mixMixStart}
\State $v_{x}\gets T\left(x_s\right)$ \Comment{code representation}\label{line:mixSelectDataXs}
\State $v_{ref}\gets T\left(x_{ref}\right)$ \Comment{code representation}\label{line:mixSelectDataXref}
\State $\lambda\gets Beta\left(\alpha\right)$ \Comment{hyperparameter}
\State $x_{mix}\gets \lambda v_{x}+\left(1-\lambda\right)v_{ref}$\Comment{data generation}
\State $y_{mix}\gets \lambda y_s+\left(1-\lambda\right)y$ \Comment{label generation}\label{line:labelGeneration}
\State $X_{mix}\gets X_{mix}\cup x_{mix}$
\State $Y_{mix}\gets Y_{mix}\cup y_{mix}$ \label{line:mixAddtoY}
\EndFor
\State $M.\mathsf{Fit}\left(X_{mix}, Y_{mix}\right)$ \Comment{training with augmented data}
\EndFor
\State \Return $M$
\end{algorithmic}
\end{algorithm}
Essentially, Algorithm~\ref{alg:mixcode} is different from the existing approaches in Algorithm~\ref{alg:basic} in the way it augments the training data in each training epoch.
\begin{figure}[h]
	\centering
	\includegraphics[width=0.9\linewidth]{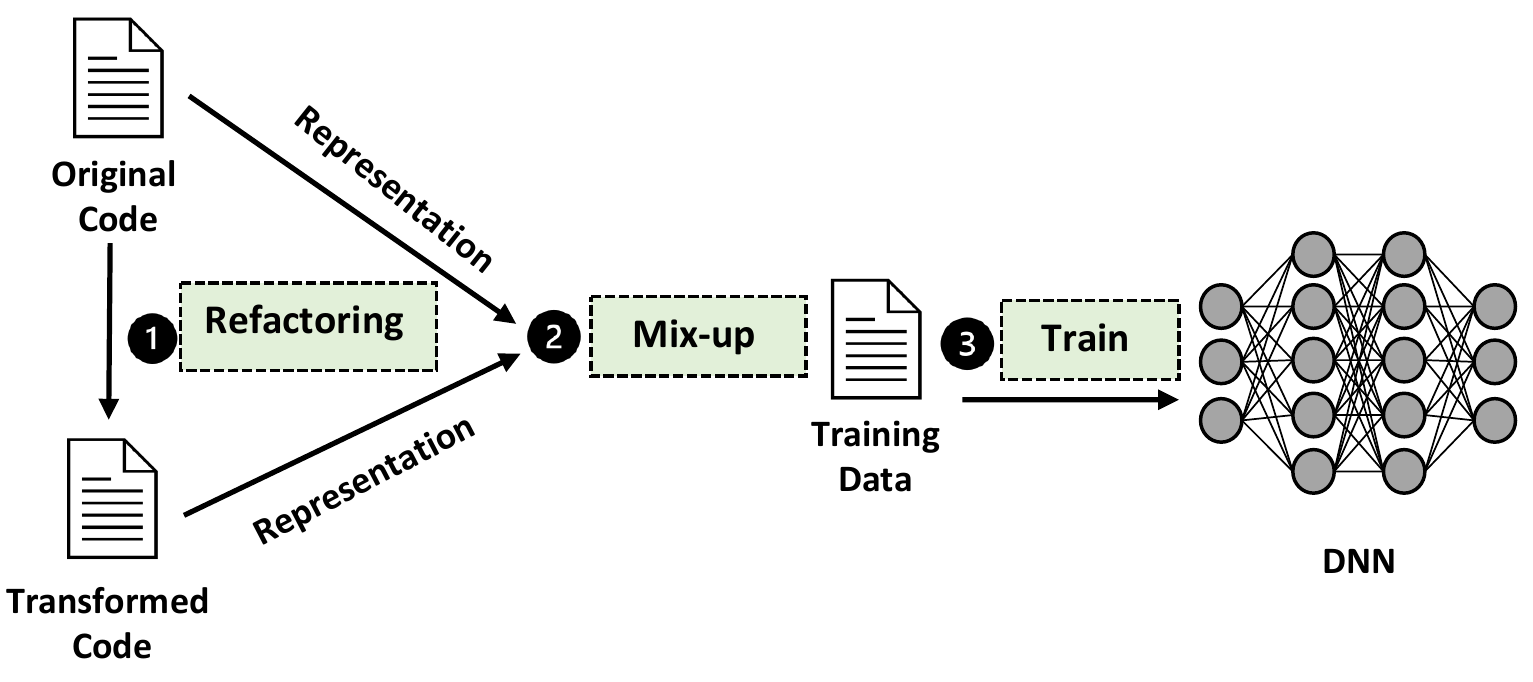}
	\caption{Workflow of \mixcode within one training epoch.}
	\label{fig:workflow_mixcode}
\end{figure}
Figure~\ref{fig:workflow_mixcode} presents an overview of \mixcode in one epoch. Concretely, this process consists of the following three phases:
\begin{enumerate}[leftmargin=*]
\item\label{phase:refactor} \mixcode loads the raw data, which consists of the \emph{original code}, from the original training set $(X, Y)$. Then, for each data, it randomly selects one refactoring method (Line~\ref{line:mixRandomSelection}) and applies it to the original code to obtain the \emph{transformed code} (Line~\ref{line:mixRefactAdd}). Code refactoring is a technique that restructures the code
without changing its semantic behaviors~\cite{kaur2016analysis,lacerda2020code}. The current version of \mixcode supports 18 different refactoring methods, which we elaborate on later in Section~\ref{sec:refactorMethods}. As a result, all the training epochs have different sets of transformed code generated by different refactoring methods. 
\item\label{phase:mixup} \mixcode randomly chooses one data from the original code and one data from the transformed code, respectively (Line~\ref{line:mixSelectDataXs} and \ref{line:mixSelectDataXref}), and then mixes these two data following Eq.~(\ref{eqn:mixup}), shown as Lines \ref{line:mixMixStart}-\ref{line:mixAddtoY} in Algorithm~\ref{alg:mixcode}. Here, $v_{x}$ and $v_{ref}$, corresponding to $x^i$ and $x^j$ in Eq.~(\ref{eqn:mixup}), are data under proper code representation. Generally, the data format is a sequence of token values. Moreover, $y_{s}$ and $y$ in Line~\ref{line:labelGeneration}, corresponding to $y^i$ and $y^j$, are the one-hot values of labels, which are the same as the original Mixup approach in Section~\ref{sec:originMixup}. In this way, we produce new data ($x_{mix}, y_{mix}$) in a similar way as the original Mixup does, and add them into the training set $(X_{mix}, Y_{mix})$.
\item\label{phase:train} Finally, the mixed data set $(X_{mix}, Y_{mix})$ is used as the training data for this epoch. Note that, although evaluated on image data in~\cite{zhang2017mixup}, the Mixup technique is not limited to the continuous representation space. The core of Mixup is to combine sample-target pairs to increase the training data size. In addition, it has been proven to be applicable to the discrete representation space for NLP tasks~\cite{guo2019augmenting}. Similarly, \mixcode supports both Integer-type and Float-type input data by simply setting the input type of embedding layers as float (e.g., \texttt{x = torch.tensor(dataset, dtype=torch.float)}).
\end{enumerate}

\noindent
\textbf{Influence factors.} In phase~\ref{phase:mixup}, we notice that two factors could affect the performance of \mixcode, namely, the candidate data pairs (namely, Line~\ref{line:mixMixStart} in Algorithm~\ref{alg:mixcode}) used for Mixup and the hyperparameter $\alpha$. For the candidate data, we have three optional combinations, 1) mixing the two original codes, 2) mixing the original code and transformed data, and 3) mixing two transformed codes. On the other side, the hyperparameter $\alpha$ controls the percentage ($\lambda$) of code content used from two parts for Mixup. As mentioned in Section~\ref{sec:background}, $\lambda$ is a value with the [0, 1] range and is sampled from a Beta distribution~\cite{mcdonald1995generalization} parameterized by $\alpha$, and $\alpha = 0.2$ is the recommended setting for image classification tasks in the original work of Mixup. In our evaluation, we study the settings of $\alpha$  from 0.05 to 0.5 to try to find a suitable setting for source code classification.

\subsection{Refactoring Methods}\label{sec:refactorMethods}
\begin{table*}[]
\centering
\caption{Description and examples of 18 types of refactoring methods.}
\label{tab:refac}
\resizebox{1.0\textwidth}{!}{
\begin{tabular}{llll}
\toprule
\textbf{No.} & \textbf{Refactoring method} & \textbf{Functionality} & \textbf{Example} \\ \hline
1 & API renaming & \begin{tabular}[c]{@{}l@{}}Rename an API by a synonym of its name. Only for token-based\\ code learning tasks.\end{tabular} &  $numpy.add\left(\right) \rightarrow numpy.delete\left(\right)$\\ \hline

2 & Arguments adding & Add an unused argument to a function definition. & $def$ $func\left(a,b\right)\rightarrow def$ $func\left(a,b,c\right)$ \\ \hline

3 & Arguments renaming & Rename an augment by a synonym of its name. & $def$ $func\left(number\right) \rightarrow def$ $func\left(size\right)$ \\ \hline

4 & Dead for adding & Add an unreachable for loop at a randomly selected location. & add: $for$ $i$ $in$ $range\left(0\right):$ $print\left(0\right)$ \\ \hline

5 & Dead if adding & \begin{tabular}[c]{@{}l@{}}Add an unreachable if statement at a randomly selected\\ location in the code.\end{tabular} & add: $if\left(1==0\right):$  $print\left(0\right)$\\ \hline

6 & Dead if else adding & \begin{tabular}[c]{@{}l@{}}Add an unreachable if-else statement at a randomly selected\\ location in the code.\end{tabular} & add: $\ print\left(0\right) if\ \left(1\ ==\ 0\right)\ else\ print\left(1\right)$\\ \hline

7 & Dead switch adding & \begin{tabular}[c]{@{}l@{}}Add an unreachable switch statement at a randomly selected \\ location.\end{tabular} & \begin{tabular}[c]{@{}l@{}}$int$ $a=0$; $switch$ $\left(a\right)$ $case$ 1:\\ $System.out.println\left(``pass''\right)$; $break$;\\ $default$: $System.out.println\left(``pass''\right)$;\end{tabular} \\ \hline

8 & Dead while adding & Add an unreachable while loop at a randomly selected location. & add: $while \left(1==0\right)$: $print\left(0\right)$\\ \hline

9 & Duplication & \begin{tabular}[c]{@{}l@{}}Duplicate a randomly selected assignment and insert it\\ to its next line.\end{tabular} & $a\ =\ 1\rightarrow a\ =\ 1;\ a\ =\ 1$ \\ \hline

10 & Filed enhancement & \begin{tabular}[c]{@{}l@{}}Enhance the rigor of the code by checking if the input of \\ each argument is None.\end{tabular} & \begin{tabular}[c]{@{}l@{}}$def$ $\left(a\right)$: $\rightarrow def$ $\left(a\right)$: \\$if$ $a== None$: $print\left(``please\ check\ your\ input.''\right)$ \end{tabular}\\ \hline

11 & For loop enhancement & \begin{tabular}[c]{@{}l@{}}Enhance the for loop conditions by complementing the \\ lower and upper bound.\end{tabular} & $for\ i\ in\ range\left(10\right)\rightarrow for\ i\ in\ range\left(0,\ 10\right)$  \\ \hline

12 & If enhancement & Change an if condition to an equivalent logic. & $if$ $True$: $\rightarrow$ $if \left(0==0\right)$ \\ \hline

13 & Local variable adding & Add an unused local variable. & add: $a=1$ \\ \hline

14 & Local variable renaming & \begin{tabular}[c]{@{}l@{}}Rename a local variable by a synonym of its name and\\ recursively update all related variables.\end{tabular} & $number\ =1\rightarrow size\ =1$ \\ \hline

15 & Method name renaming & Rename a method by a synonym of its name. & $def$  $count\left(a\right)\rightarrow def$ $compute\left(a\right)$  \\ \hline

16 & Plus zero & \begin{tabular}[c]{@{}l@{}}Select an numerical assignment of mathematical calculation\\ and plus zero to its value.\end{tabular} & $a=1\rightarrow a=1+0$ \\ \hline

17 & Print adding & Add a print line at a randomly selected location. & add: $print\left(1\right)$ \\ \hline

18 & Return optimal & Change the return content to a variant with the same effect. & $return$ 1 $\rightarrow return$ 0 $if \left(1==0\right)$ $else$ 1 \\ \bottomrule

\end{tabular}
}
\end{table*}

\begin{figure}[!tb]
	\centering
	\includegraphics[width=\linewidth]{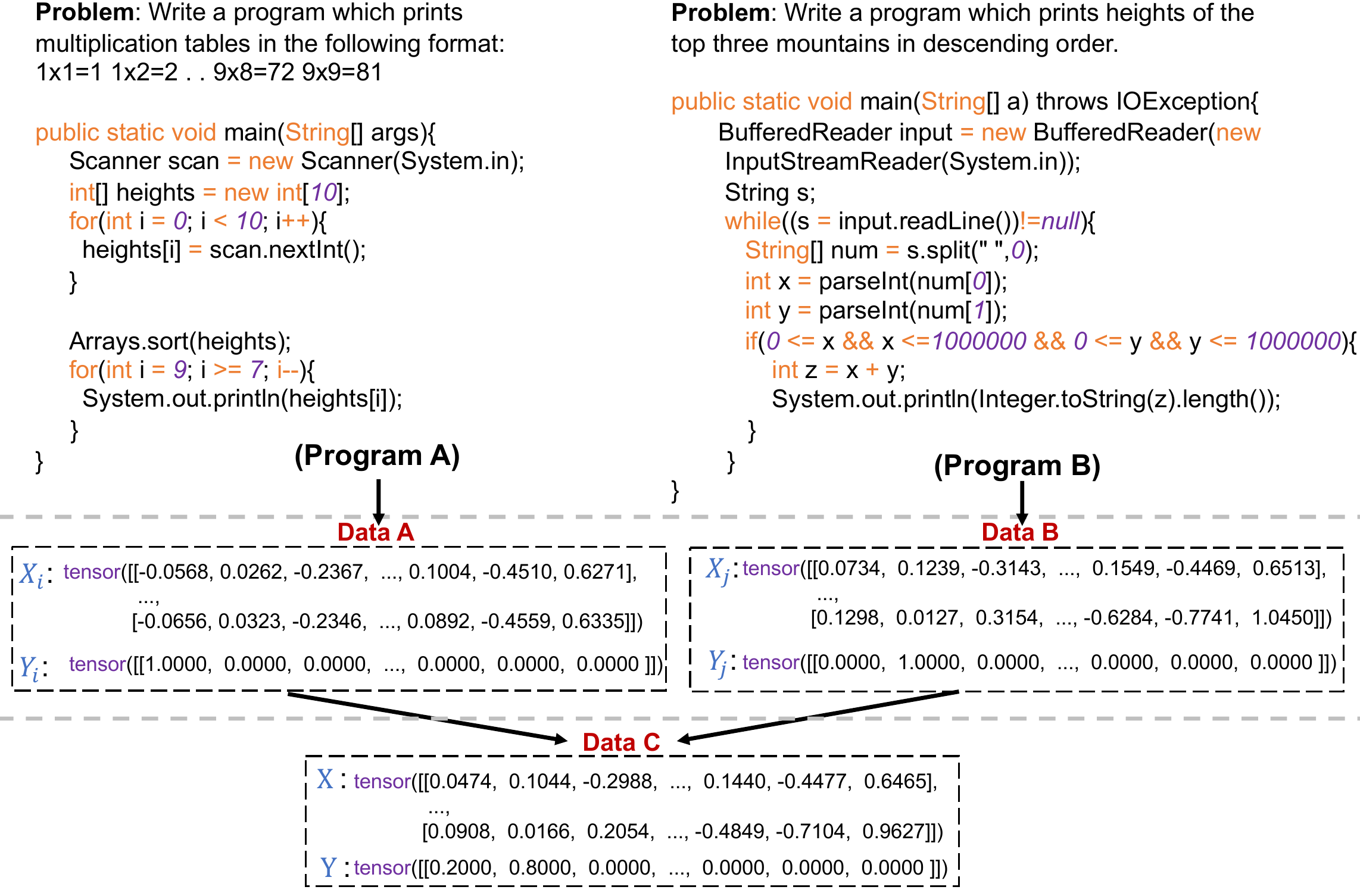}
	\caption{An example of two programs mixed by \mixcode.}
	\label{fig:example_mixcode}
\end{figure}

As we introduced, 
code refactoring is a technique that restructures the code while keeping its semantic behaviors~\cite{kaur2016analysis,lacerda2020code}. The original purpose of code refactoring is to improve readability and reduce code complexity. Thus, programmers can clean up complicated code and reduce technical debt. Meanwhile, refactoring makes it easier for developers to maintain and add new features to the clean code, which is an important step in software maintenance. Typically, refactoring makes a small change in source code that preserves the behavior of the program based on a series of standardized micro-modifies. Several refactoring methods have been proposed and studied, such as replacing a variable, modifying (including adding or simplifying conditional expressions and method calls), moving features between objects, and organizing data. 

In the first step of \mixcode, in addition to the original code, we utilize multiple code refactoring methods to generate more diverse code as the candidate training data. \mixcode supports 18 types of refactoring methods from the literature~\cite{wei2021cocofuzzing, pour2021search}. The functionality of each method and a corresponding example are listed in Table~\ref{tab:refac}. Note that some transformations may alter code semantics, but not in a way that can affect model decisions, i.e., all transformations are label-preserving. Existing code learning models mainly use two types of code representation: token sequence and abstract syntax tree (AST)~\cite{le2020rep}. All the \mixcode-supported refactoring methods except API Renaming can be applied to these two prepossessing formats, and therefore, \mixcode is equipped with strong flexibility.
\begin{wrapfigure}[12]{c}{0.55\linewidth}
 \centering
 \includegraphics[width =\linewidth]{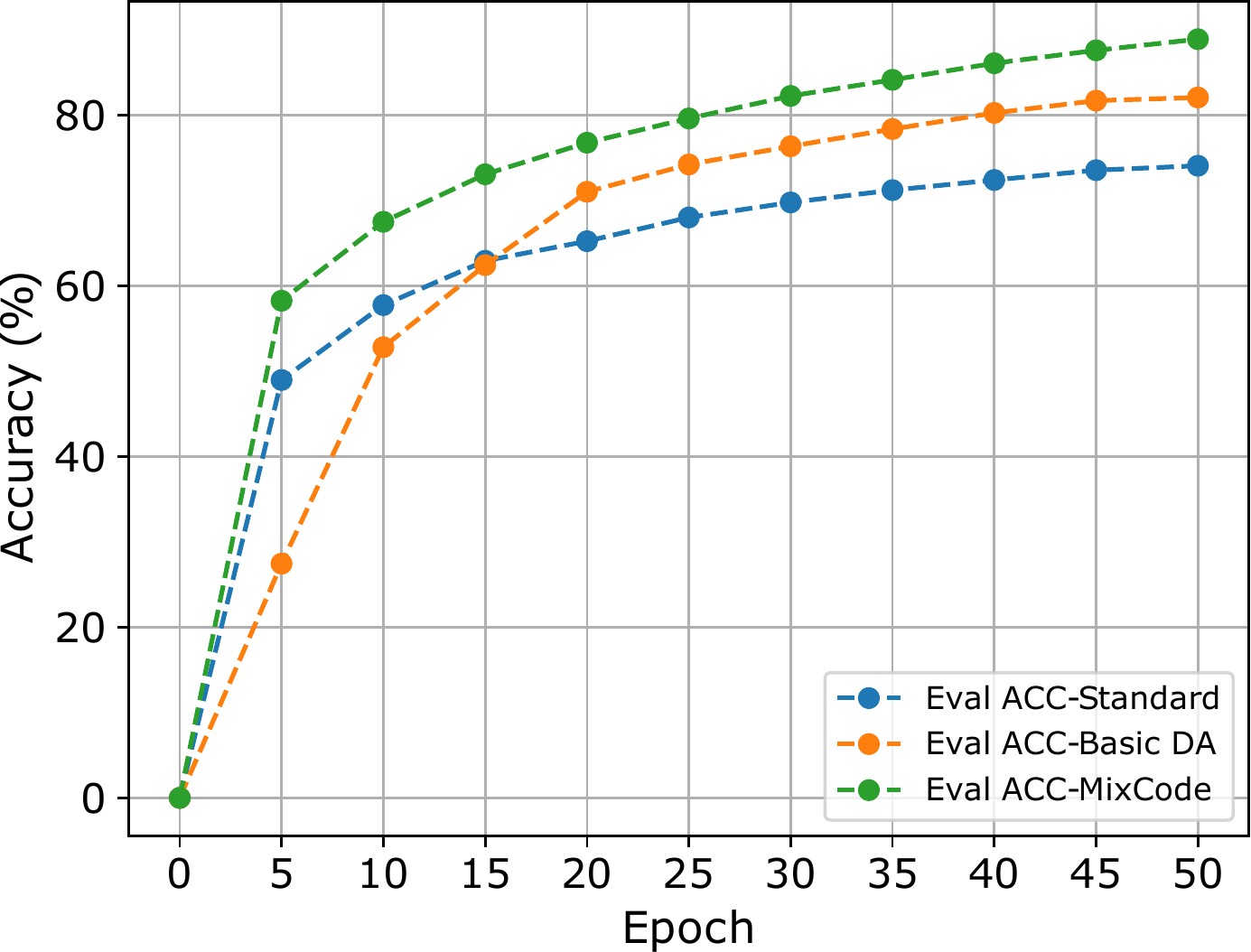}
 \caption{Test accuracy of models trained by using different methods (dataset: Java250, model: BagofToken).}\label{fig:example_mixcode_figure}
\end{wrapfigure}
\subsection{A Case Study}
To better understand how \mixcode works, we use an example to show its workflow, as depicted in Figure~\ref{fig:example_mixcode}. We first transform the source code (Program A and B from the JAVA250 dataset with label 0 and label 1, respectively) into vectors through CodeBERT~\cite{feng2020codebert} and the labels into one-hot vectors (see Data A and Data B in Figure~\ref{fig:example_mixcode}). $X$ is the value of each token, and $Y$ is the one-hot format label (250 classes in total). Next, we linearly mix the code and label vectors, respectively, of these two programs as the final input to train the model (See Data C in Figure~\ref{fig:example_mixcode}). In this example, we set $\lambda$ in Eq.~(\ref{eqn:mixup}) as 0.2 and perform the input mixing. 

Then, we use one example to show the training process of using different training methods, as shown in Figure \ref{fig:example_mixcode_figure}. We can see that \mixcode leads to faster model convergence than the other two methods, which draws a similar conclusion to the usage of Mixup (and its variants) in other fields~\cite{hendrycks2019augmix, hendrycks2022pixmix,zhang2020does}.


\section{Experimental Setup}
\label{sec:setup}
To evaluate the effectiveness of \mixcode, we conduct experiments on two popular programming languages (Java and Python), two important downstream tasks (problem classification and bug detection), and seven DNN model architectures, including two pre-trained model CodeBERT and GraphCodeBERT. Table~\ref{tab:data_model} presents the details of the datasets and models used in the evaluation. 

\begin{table}[]
\caption{Details of datasets and DNNs. \#Training, \#Ori Test, and \#Robust Test represent the number of training data, original test data, and test data for generalization evaluation, respectively.}
\label{tab:data_model}
\resizebox{\columnwidth}{!}{
\begin{tabular}{ccccccc}
\toprule
\textbf{Dataset} & \textbf{Language} & \textbf{Task} & \textbf{\#Training} & \textbf{\#Ori Test} & \multicolumn{1}{l}{\textbf{\#Robust Test}} & \textbf{Model} \\ \hline
\multirow{2}{*}{\textbf{JAVA250}} & \multirow{2}{*}{Java} & \multirow{2}{*}{\begin{tabular}[c]{@{}c@{}}Problem \\ classification\end{tabular}} & \multirow{2}{*}{48000} & \multirow{2}{*}{15000} & \multirow{2}{*}{75000} & \multirow{4}{*}{\begin{tabular}[c]{@{}c@{}}BagofToken\\ SeqofToken\\ CodeBERT\\ GraphCodeBERT\end{tabular}} \\
 &  &  &  &  &  &  \\ \cline{1-6}
\multirow{2}{*}{\textbf{Python800}} & \multirow{2}{*}{Python} & \multirow{2}{*}{\begin{tabular}[c]{@{}c@{}}Problem \\ classification\end{tabular}} & \multirow{2}{*}{153600} & \multirow{2}{*}{48000} & \multirow{2}{*}{240000} &  \\
 &  &  &  &  &  &  \\ \hline
\multirow{3}{*}{\textbf{CodRep1}} & \multirow{3}{*}{Java} & \multirow{3}{*}{\begin{tabular}[c]{@{}c@{}}Bug \\ detection\end{tabular}} & \multirow{3}{*}{6944} & \multirow{3}{*}{772} & \multirow{3}{*}{7716} & \multirow{6}{*}{\begin{tabular}[c]{@{}c@{}}GCN\\ GAT\\ GGNN \\ CodeBERT\\ GraphCodeBERT\end{tabular}} \\
 &  &  &  &  &  &  \\
 &  &  &  &  &  &  \\ \cline{1-6}
\multirow{3}{*}{\textbf{Refactory}} & \multirow{3}{*}{Python} & \multirow{3}{*}{\begin{tabular}[c]{@{}c@{}}Bug \\ detection\end{tabular}} & \multirow{3}{*}{3380} & \multirow{3}{*}{423} & \multirow{3}{*}{4225} &  \\
 &  &  &  &  &  &  \\
 &  &  &  &  &  &  \\ \bottomrule
\end{tabular}
}
\end{table}

\noindent
\textbf{Task and Dataset.} \mixcode is suitable for all code classification problems. We select two widely studied tasks, problem classification and bug detection, in our evaluation to demonstrate the effectiveness of \mixcode. \textit{Problem classification} is a basic code learning task for collecting the target code from code pools with massive source code (e.g., GitHub). Given a number of problem descriptions and the candidate source code, the model will predict the problem that this code is trying to solve. For this task, our experiments use two recently released datasets, JAVA250 and Python800~\cite{puri2021codenet}. JAVA250 is used for Java program classification with 250 classification problems, and each problem includes 300 Java programs. It also provides two different program representations, which are BagofToken and SeqofToken. Python800 is a dataset for Python program classification, including 800 classification problems with 300 Python programs in each problem. BagofToken and SeqofToken are also available in Python800. \textit{Bug detection} tries to identify whether one code has bugs or not. Generally, bug detection can be seen as a binary classification problem. Preparing bug detection datasets is difficult since it requires a pair of codes with and without bugs. Crawling two versions of code before and after commits from GitHub is the common way to collect such pairs. However, human effort is required to check if the commit is to solve a bug manually. For this task, we use two open datasets, Refactory~\cite{hu2019re} and CodRep1~\cite{zhong2015empirical} in our study. Refactory includes 2,242 correct and 1,783 buggy Python programs written by real-world undergraduate students. CodRep1 provides 3,858 program pairs (buggy program and its fixed version) from real bug fixes. We use each dataset's originally provided training data for the model training.

\noindent
\textbf{Model.} To avoid the model-depended issue, we build at least four model architectures for each dataset. In problem classification tasks, we follow the recommendation of~\cite{puri2021codenet} and build a bag of token-based FNN (BagofToken) and a sequence of token-based CNN (SeqofToken) for each dataset. FNN is a basic DNN type with only dense layers, while CNN is the advanced model type with convolutional layers and achieves great success in image classification tasks. To be simplified, in the remaining part, we use SeqofToken to represent the sequence of token-based CNN, which represents the code with its order of tokens, and BagofToken to represent the bag of token-based FNN, which represents a code sample with the relative frequencies of operator and keyword token occurrences, respectively. In the bug detection tasks, we follow the work~\cite{zhou2019devign} and build three types of GNN models, Gated Graph Sequence Neural Networks (GGNNs)~\cite{li2015gated}, Graph Convolutional Networks (GCNs)~\cite{kipf2016semi}, and Graph Attention Networks (GATs)~\cite{velickovic2017graph}. GGNN is a graph neural network that updates the new hidden state with Gated Recurrent Unit (GRU).  GCN is a variant of convolution neural networks that operate on graph-structured data. GAT mainly applies the attention mechanism to graph message passing. A code graph is an inter-procedural graph of program instructions where edges represent data flow and control flow information. In graph-based code models,  we mix the graph data vectors of code representation, which are produced by the GNN using the node and edge embeddings, not code directly. Besides, we also include CodeBERT~\cite{feng2020codebert} and GraphCodeBERT~\cite{guo2020graphcodebert} in both of the code tasks, as they have demonstrated state-of-the-art performance in a variety of code processing tasks~\cite{lu2021codexglue,zhou2021assessing}. CodeBERT is a bimodal pre-trained model that takes as input natural language text and code data and produces the general representations of the text and code. GraphCodeBERT is a pre-trained model that represents the code using data-flow information to consider the semantic-level structure of code. 

\noindent
\textbf{Baseline.} We compare \mixcode with two baselines. The first one is the basic data augmentation method, in which we train the model while conducting random code refactoring on the training data at each training epoch. This data augmentation method is used in existing works~\cite{pour2021search, yefet2020adversarial}. The second baseline is the standard training process without any data augmentation.

\noindent
\textbf{Evaluation measures.} For the model testing, we evaluate the test accuracy and robustness of DNNs. The test accuracy is the percentage of correctly classified data over the entire test data. For the robustness evaluation, we first generate new test sets by the refactoring methods from the original test set, then calculate the percentage of correctly classified data from this new set. The robustness reflects the generalization ability of the trained model, i.e., the model performance when facing more diverse unseen data. 

\noindent
\textbf{Implementation and Environments.} The pure Python language implements the core code of \mixcode with only the Numpy package. That means \mixcode can be easily reused in any DL framework for other classification tasks. We provide two versions of code refactoring methods that support both Java and Python languages. The models of the classification tasks are built using TensorFlow2.3 and Keras2.4.3 frameworks, and the models of the bug detection tasks are built using PyTorch1.6.0. For the SeqOfToken, BagOfToken, CodeBERT, and GraphCodeBERT experiments, we set the training epoch configuration to 50 and the GNNs experiment to 100. We conduct our experiments on an NVIDIA Tesla V100 16G SXM2 GPU. We train each model 5 times to reduce the effect of randomness and report the average results with standard deviation. 

\section{Results Analysis}
\label{sec:results}

\subsection{RQ1: Effectiveness of \mixcode}
The upper component in Table~\ref{tab:rq1_acc} presents the test accuracy of trained models on original test data. The first conclusion to be drawn from the result is that \mixcode almost always outperforms the two baselines regardless of datasets and models, showing that data augmentation is useful to improve the model performance compared to standard training. However, simply adding data into the training set only slightly improves the basic strategy, especially in bug detection tasks (CodRep1 and Refactory). For example, in Refactory with GAT, the basic data augmentation only improves the accuracy by up to 0.29\%. This phenomenon is similar to the findings in~\cite{yefet2020adversarial}, which reveal that traditional adversarial training (simply adding adversarial examples to the training data) is not sufficient to improve the robustness of source code models. By contrast, \mixcode has a significant improvement (up to 5.69\%) compared to the basic data augmentation. In bug detection (CodRep1 and Refactory), where the basic data augmentation is not helpful, \mixcode can increase the model accuracy with an impressive improvement by up to 6.24\%.

\begin{table}[]
\caption{Effectiveness of \mixcode concerning test accuracy and robustness (average$\pm$ standard deviation, \%) on original test data and transformed test data. Standard: standard training without data augmentation. Basic: basic data augmentation. The value with a gray background indicates the best result, and the value highlighted in red shows the accuracy difference between the best and the second-best. The higher, the better.}
\label{tab:rq1_acc}
\centering
\resizebox{0.9\columnwidth}{!}{
\begin{tabular}{llccc}
\toprule
\textbf{Dataset} & \textbf{DNN} & \textbf{Standard} & \textbf{Basic} & \textbf{\mixcode} \\ \hline
\multicolumn{5}{c}{\textbf{Test Accuracy}} \\ \hline
 & \textbf{BagofToken} & 71.66$\pm$0.03 & 76.63$\pm$0.08 & \cellcolor[HTML]{C0C0C0} 82.32$\pm$0.07 ({\color{red} 5.69$\uparrow$}) \\
\multirow{-0.5}{*}{\textbf{JAVA250}} & \textbf{SeqofToken} & 86.57$\pm$0.06 & 93.92$\pm$0.15 & \cellcolor[HTML]{C0C0C0} 94.52$\pm$0.23 ({\color{red}0.60$\uparrow$}) \\ 
& \textbf{CodeBERT} & 96.37$\pm$0.02 & 96.46$\pm$0.07 & \cellcolor[HTML]{C0C0C0} 96.98$\pm$0.09 ({\color{red}0.52$\uparrow$}) \\
& \textbf{GraphCodeBERT} &96.48$\pm$0.03 & 96.51$\pm$0.06 & \cellcolor[HTML]{C0C0C0} 97.16$\pm$0.12 ({\color{red}0.65$\uparrow$})\\\hline

 & \textbf{GGNN} & 62.69$\pm$0.32 & 63.07$\pm$0.35 & \cellcolor[HTML]{C0C0C0}65.42$\pm$0.35 ({\color{red}2.35$\uparrow$}) \\
 & \textbf{GCN} & 61.38$\pm$0.44 & 62.68$\pm$0.56 & \cellcolor[HTML]{C0C0C0}64.18$\pm$0.44 ({\color{red}1.50$\uparrow$}) \\
\multirow{-1}{*}{\textbf{CodRep1}} & \textbf{GAT} & 61.27$\pm$0.26 & 62.07$\pm$0.33 & \cellcolor[HTML]{C0C0C0}64.09$\pm$0.36 ({\color{red}2.02$\uparrow$}) \\ 

& \textbf{CodeBERT} & 69.12$\pm$0.03 & 71.22$\pm$0.02 & \cellcolor[HTML]{C0C0C0}72.96$\pm$0.05 ({\color{red}1.74$\uparrow$}) \\ 
& \textbf{GraphCodeBERT} & 70.05$\pm$0.11 & 71.35$\pm$0.09 & \cellcolor[HTML]{C0C0C0}73.34$\pm$0.13 ({\color{red}1.99$\uparrow$}) \\\hline

 & \textbf{BagofToken} & 67.31$\pm$0.05 & 67.46$\pm$0.08 & \cellcolor[HTML]{C0C0C0}68.27$\pm$0.08 ({\color{red}0.81$\uparrow$}) \\
 \multirow{-0.5}{*}{\textbf{Python800}} & \textbf{SeqofToken} & 82.65$\pm$0.32 & 83.26$\pm$0.41 & \cellcolor[HTML]{C0C0C0}85.00$\pm$0.20 ({\color{red}1.74$\uparrow$}) \\ 
 
    & \textbf{CodeBERT} & 96.05$\pm$0.02 & 96.16$\pm$0.01 & \cellcolor[HTML]{C0C0C0}96.79$\pm$0.06 ({\color{red}0.63$\uparrow$}) \\ 
    
    & \textbf{GraphCodeBERT} & 96.27$\pm$0.05 & 96.29$\pm$0.03 & \cellcolor[HTML]{C0C0C0}97.09$\pm$0.08 ({\color{red}0.80$\uparrow$}) \\\hline
    
 & \textbf{GGNN} & 81.96$\pm$0.22 & 81.98$\pm$0.23 & \cellcolor[HTML]{C0C0C0}88.22$\pm$0.18 ({\color{red}6.24$\uparrow$}) \\
 & \textbf{GCN} & 80.12$\pm$0.25 & 80.23$\pm$0.37 & \cellcolor[HTML]{C0C0C0}84.16$\pm$0.25 ({\color{red}3.93$\uparrow$}) \\
\multirow{-1}{*}{\textbf{Refactory}} & \textbf{GAT} & 79.76$\pm$0.19 & 80.05$\pm$0.28 & \cellcolor[HTML]{C0C0C0}83.38$\pm$0.31 ({\color{red}3.33$\uparrow$}) \\ 

& \textbf{CodeBERT} & 95.88$\pm$0.42 & 96.09$\pm$0.33 & \cellcolor[HTML]{C0C0C0}97.57$\pm$0.37 ({\color{red}1.48$\uparrow$}) \\ 
& \textbf{GraphCodeBERT} & 96.81$\pm$0.17 & 96.92$\pm$0.11 & \cellcolor[HTML]{C0C0C0}98.16$\pm$0.21  ({\color{red}1.24$\uparrow$}) \\\hline

\multicolumn{5}{c}{\textbf{Robustness}} \\ \hline
 & \textbf{BagofToken} & 40.21$\pm$0.09 & 52.11$\pm$0.06 & \cellcolor[HTML]{C0C0C0}78.17$\pm$0.05 ({\color{red}26.06$\uparrow$}) \\
 \multirow{-0.5}{*}{\textbf{JAVA250}} & \textbf{SeqofToken} & 57.83$\pm$0.01 & 84.94$\pm$0.02 & \cellcolor[HTML]{C0C0C0}85.60$\pm$0.01 ({\color{red}0.66$\uparrow$}) \\ 
 
 &\textbf{CodeBERT} & 89.71$\pm$0.09 & 92.19$\pm$0.05 & \cellcolor[HTML]{C0C0C0}93.17$\pm$0.11 ({\color{red}0.98$\uparrow$}) \\ 
 &\textbf{GraphCodeBERT} & 90.56$\pm$0.03 & 92.78$\pm$0.07 & \cellcolor[HTML]{C0C0C0}94.07$\pm$0.16 ({\color{red}1.29$\uparrow$}) \\\hline

 & \textbf{GGNN} & 38.84$\pm$0.02 & 50.76$\pm$0.03 & \cellcolor[HTML]{C0C0C0}55.01$\pm$0.01 ({\color{red}4.25$\uparrow$}) \\
 & \textbf{GCN} & 38.40$\pm$0.01 & 49.97$\pm$0.02 & \cellcolor[HTML]{C0C0C0}54.03$\pm$0.03 ({\color{red}4.06$\uparrow$}) \\
\multirow{-1}{*}{\textbf{CodRep1}} & \textbf{GAT} & 38.07$\pm$0.02 & 49.46$\pm$0.04 & \cellcolor[HTML]{C0C0C0}53.81$\pm$0.03 ({\color{red}4.35$\uparrow$}) \\

 & \textbf{CodeBERT} & 61.11$\pm$0.04 & 62.28$\pm$0.03 & \cellcolor[HTML]{C0C0C0}66.15$\pm$0.09 ({\color{red}3.87$\uparrow$}) \\
  & \textbf{GraphCodeBERT} & 61.03$\pm$0.11 & 61.86$\pm$0.04 & \cellcolor[HTML]{C0C0C0}65.91$\pm$0.13 ({\color{red}4.05$\uparrow$}) \\ \hline

 & \textbf{BagofToken} & 38.76$\pm$0.02 & 39.04$\pm$0.02 & \cellcolor[HTML]{C0C0C0}63.65$\pm$0.02 ({\color{red}24.61$\uparrow$}) \\
  \multirow{-0.5}{*}{\textbf{Python800}} & \textbf{SeqofToken} & 58.18$\pm$0.03 & 80.81$\pm$0.02 & \cellcolor[HTML]{C0C0C0}82.94$\pm$0.02 ({\color{red}2.13$\uparrow$}) \\

  & \textbf{CodeBERT} & 89.64$\pm$0.04 & 89.97$\pm$0.01 & \cellcolor[HTML]{C0C0C0}92.16$\pm$0.08 ({\color{red}2.19$\uparrow$}) \\
  & \textbf{GraphCodeBERT} & 91.01$\pm$0.02 & 91.85$\pm$0.04 & \cellcolor[HTML]{C0C0C0}94.56$\pm$0.07 ({\color{red}2.71$\uparrow$}) \\\hline

 & \textbf{GGNN} & 66.37$\pm$0.02 & 67.34$\pm$0.02 & \cellcolor[HTML]{C0C0C0}72.81$\pm$0.03 ({\color{red}5.47$\uparrow$}) \\
 & \textbf{GCN} & 64.07$\pm$0.01 & 64.32$\pm$0.03 & \cellcolor[HTML]{C0C0C0}67.73$\pm$0.03 ({\color{red}3.41$\uparrow$}) \\
\multirow{-1}{*}{\textbf{Refactory}} & \textbf{GAT} & 62.03$\pm$0.02 & 62.31$\pm$0.01 & \cellcolor[HTML]{C0C0C0}66.49$\pm$0.01 ({\color{red}4.38$\uparrow$}) \\ 

& \textbf{CodeBERT} & 92.14$\pm$0.12 & 92.54$\pm$0.09 & \cellcolor[HTML]{C0C0C0}95.41$\pm$0.06 ({\color{red}2.87$\uparrow$}) \\ 
& \textbf{GraphCodeBERT} & 92.02$\pm$0.05 & 92.15$\pm$0.03 & \cellcolor[HTML]{C0C0C0}95.23$\pm$0.11 ({\color{red}3.08$\uparrow$}) \\ \hline
\end{tabular}
}
\vspace{-5mm}
\end{table}
The lower part in Table~\ref{tab:rq1_acc} presents the robustness of different trained models on the transformed test data. First of all, from the test accuracy to the robustness, we observe that the results of all standard-trained models drop significantly, e.g., from 71\% to 40\% for JAVA250-BagofToken. This phenomenon confirms that only using original training data to train the model will result in a bad generalization property, i.e., the model can not handle the unseen data even if the data is functionally the same as the training data. However, the drop is reduced via data augmentation, especially by \mixcode. \mixcode achieves the best results in all cases, and the basic data augmentation usually performs similarly to the standard training. More specifically, the results show that it outperforms the basic data augmentation by up to 26.06\%, and on average, 5.58\%, which is remarkable. Particularly, in the problem classification task (JAVA250 and Python800), all the models have more than 20\% accuracy improvement. And perhaps surprisingly, the robustness is already close to the original test accuracy, e.g., Python800-BagofToken, original accuracy 68.27\% vs. robustness 63.65\%. In the bug detection task (CodRep1 and Refactory), although the improvement is lower than in the problem classification (JAVA250 and Python800), it is still promising (from 2.87\% to 5.47\%).

\vspace{1mm}

\noindent\colorbox{gray!20}{\framebox{\parbox{0.96\linewidth}{
\textbf{Answer to RQ1}: \mixcode is effective in enhancing model performance. It outperforms the basic data augmentation by up to 6.24\% and 26.06\% original test accuracy and robustness improvement, respectively.}}}

\subsection{RQ2: Impact of Mixup Strategies}
\label{seq:rq2_2}
In the default setting of \mixcode in step \ref{phase:mixup}, a pair of randomly selected original code and transformed code is mixed (\emph{Ori+Ref}). To investigate the usefulness of this mixing strategy, we compare (\emph{Ori+Ref}) to other two types of Mixup strategies, original code mixing original code (\emph{Ori+Ori}) and transformed code mixing transformed code (\emph{Ref+Ref}). 

Table~\ref{tab:rq2_acc} shows the effectiveness comparison of these three strategies. First, compared to the results by producing standard training in Table~\ref{tab:rq1_acc}, we can see that all three strategies benefit the model performance. Meanwhile, in most cases ( 84 out of 108), the Mixup strategy outperforms the basic data augmentation. Next, we compare the effectiveness of different Mixup strategies. Considering the test accuracy, the results show that for the Java language (JAVA250, CodRep1), \textit{Ori+Ref} consistently outperforms the other two combinations on all the tasks. \textit{Ref+Ref}, the second-best, is slightly better than \textit{Ori+Ori}. However, overall, the difference between these three strategies is small, e.g., 94.49\% vs 94.52\%, 93.18\%. For the Python language (Python800, Refactory), we can find that there is no one can always be the best. But the \textit{Ori+Ref} is still in the top-2 places in all cases. And even if \textit{Ori+Ref} is in the second place, the gap between it and the best is slight (e.g., 84.12\% vs 84.61\%). Unlike Java tasks, the difference between these three combinations becomes bigger in Python Tasks, e.g., 82.46\% vs 88.22\% vs 86.58\%. Thus, if we only consider the test accuracy, \textit{Ori+Ref} is the recommended combination for \mixcode. Considering the robustness, $Ori+Ref$ consistently and significantly outperforms the other two combinations with an average of 2.62\%  better robustness than the second best. These results recommend that using original data and transformed data is a better choice for \mixcode to train robust models.  

\begin{table}[]
\centering
\caption{Comparison of different Mixup strategies on test accuracy and robustness (average$\pm$ standard deviation, \%). Ref: transformed code, Ori: original code. The value with a gray background indicates the best result. The higher, the better.}
\resizebox{0.9\columnwidth}{!}{
\label{tab:rq2_acc}
\centering
\begin{tabular}{llccc}
\toprule
& & \multicolumn{3}{c}{\textbf{Mixup Strategy}} \\ \cline{3-5}
\multirow{-2}{*}{\textbf{Dataset}} & \multirow{-2}{*}{\textbf{DNN}} & \textbf{Ori+Ori} & \textbf{Ori+Ref} & \textbf{Ref+Ref} \\ \hline
\multicolumn{5}{c}{\textbf{Test Accuracy}} \\ \hline
 & \textbf{BagofToken} & 73.09$\pm$0.03 & \cellcolor[HTML]{C0C0C0}82.32$\pm$0.07 & 82.13$\pm$0.03 \\
 \multirow{-0.5}{*}{\textbf{JAVA250}} & \textbf{SeqofToken} & 94.49$\pm$0.08 & \cellcolor[HTML]{C0C0C0}94.52$\pm$0.23 & 93.18$\pm$0.15 \\ 
 
 & \textbf{CodeBERT} & 96.54$\pm$0.02 & \cellcolor[HTML]{C0C0C0}96.98$\pm$0.09 & 95.17$\pm$0.16 \\ 
  & \textbf{GraphCodeBERT} & 96.76$\pm$0.04 & \cellcolor[HTML]{C0C0C0}97.16$\pm$0.12 & 95.56$\pm$0.19 \\ 
 \hline

 & \textbf{GGNN} & 64.37$\pm$0.25 & \cellcolor[HTML]{C0C0C0}65.42$\pm$0.35 & 65.38$\pm$0.33 \\
 & \textbf{GCN} & 63.42$\pm$0.26 & \cellcolor[HTML]{C0C0C0}64.18$\pm$0.44 & 63.89$\pm$0.27 \\
\multirow{-1}{*}{\textbf{CodRep1}} & \textbf{GAT} & 63.29$\pm$0.47 & \cellcolor[HTML]{C0C0C0}64.09$\pm$0.36 & 63.78$\pm$0.35 \\ 

& \textbf{CodeBERT} & 72.77$\pm$0.07 & \cellcolor[HTML]{C0C0C0}72.96$\pm$0.05 & 71.14$\pm$0.03 \\ 
& \textbf{GraphCodeBERT} & 73.21$\pm$0.11 & \cellcolor[HTML]{C0C0C0}73.34$\pm$0.13 & 71.78$\pm$0.16 \\\hline

 & \textbf{BagofToken} & \cellcolor[HTML]{C0C0C0}68.27$\pm$0.08 & 67.88$\pm$0.07 & 65.67$\pm$0.09 \\
 \multirow{-0.5}{*}{\textbf{Python800}} & \textbf{SeqofToken} & 84.68$\pm$0.04 & \cellcolor[HTML]{C0C0C0}85.00$\pm$0.20 & 81.22$\pm$0.45 \\ 
 
  & \textbf{CodeBERT} & 96.35$\pm$0.04 & \cellcolor[HTML]{C0C0C0}96.79$\pm$0.06 & 95.11$\pm$0.18 \\
  & \textbf{GraphCodeBERT} & 96.87$\pm$0.05 & \cellcolor[HTML]{C0C0C0}97.09$\pm$0.08 & 95.16$\pm$0.21 \\\hline
  
 & \textbf{GGNN} & 82.46$\pm$0.28 & \cellcolor[HTML]{C0C0C0}88.22$\pm$0.18 & 86.58$\pm$0.22 \\ 
& \textbf{GCN} & 81.71$\pm$0.24 & 84.12$\pm$0.17 & \cellcolor[HTML]{C0C0C0}84.61$\pm$0.25 \\
 \multirow{-1}{*}{\textbf{Refactory}} & \textbf{GAT} & 80.22$\pm$0.27 & 82.68$\pm$0.12 & \cellcolor[HTML]{C0C0C0}83.38$\pm$0.31 \\ 
 
 & \textbf{CodeBERT} & 96.98$\pm$0.41 & \cellcolor[HTML]{C0C0C0}97.57$\pm$0.37 & 95.39$\pm$0.39 \\
  & \textbf{GraphCodeBERT} & 97.78$\pm$0.27 & \cellcolor[HTML]{C0C0C0}98.16$\pm$0.21 & 96.03$\pm$0.19 \\\hline
 
 \multicolumn{5}{c}{\textbf{Robustness}} \\ \hline
 & \textbf{BagofToken} & 43.87$\pm$0.03 & \cellcolor[HTML]{C0C0C0}78.17$\pm$0.05 & 66.62$\pm$0.03 \\
 \multirow{-0.5}{*}{\textbf{JAVA250}} & \textbf{SeqofToken} & 78.27$\pm$0.02 & \cellcolor[HTML]{C0C0C0}85.60$\pm$0.01 & 84.30$\pm$0.02 \\ 
 
 & \textbf{CodeBERT} & 91.17$\pm$0.22 & \cellcolor[HTML]{C0C0C0}93.17$\pm$0.11 & 93.01$\pm$0.13 \\ 
 & \textbf{GraphCodeBERT} & 92.01$\pm$0.19 & \cellcolor[HTML]{C0C0C0}94.07$\pm$0.16 & 93.56$\pm$0.17 \\\hline  
 
 & \textbf{GGNN} & 42.40$\pm$0.02 & \cellcolor[HTML]{C0C0C0}55.01$\pm$0.01 & 49.53$\pm$0.02 \\
 & \textbf{GCN} & 42.08$\pm$0.02 & \cellcolor[HTML]{C0C0C0}54.03$\pm$0.03 & 49.42$\pm$0.03 \\
\multirow{-1}{*}{\textbf{CodRep1}} & \textbf{GAT} & 41.66$\pm$0.02 & \cellcolor[HTML]{C0C0C0}53.81$\pm$0.03 & 48.78$\pm$0.02 \\ 

& \textbf{CodeBERT} & 63.26$\pm$0.06 & \cellcolor[HTML]{C0C0C0}66.15$\pm$0.09 & 65.95$\pm$0.03 \\  
& \textbf{GraphCodeBERT} & 63.17$\pm$0.08 & \cellcolor[HTML]{C0C0C0}65.91$\pm$0.13 & 64.76$\pm$0.16 \\ \hline

 & \textbf{BagofToken} & 40.46$\pm$0.01 & \cellcolor[HTML]{C0C0C0}63.65$\pm$0.02 & 62.35$\pm$0.03 \\
 \multirow{-0.5}{*}{\textbf{Python800}} & \textbf{SeqofToken} & 75.15$\pm$0.02 & \cellcolor[HTML]{C0C0C0}82.94$\pm$0.02 & 78.54$\pm$0.01 \\ 
 
 & \textbf{CodeBERT} & 90.08$\pm$0.05 & \cellcolor[HTML]{C0C0C0}92.16$\pm$0.08 & 92.08$\pm$0.03 \\ 
 & \textbf{GraphCodeBERT} & 91.87$\pm$0.08 & \cellcolor[HTML]{C0C0C0}94.56$\pm$0.07 & 93.12$\pm$0.09 \\\hline
 
 & \textbf{GGNN} & 68.09$\pm$0.03 & \cellcolor[HTML]{C0C0C0}72.81$\pm$0.03 & 69.69$\pm$0.02 \\
 & \textbf{GCN} & 65.68$\pm$0.02 & \cellcolor[HTML]{C0C0C0}67.73$\pm$0.03 & 66.36$\pm$0.01 \\
\multirow{-1}{*}{\textbf{Refactory}} & \textbf{GAT} & 62.37$\pm$0.02 & \cellcolor[HTML]{C0C0C0}66.49$\pm$0.01 & 65.09$\pm$0.02 \\ 

& \textbf{CodeBERT} & 92.67$\pm$0.06 & \cellcolor[HTML]{C0C0C0}95.41$\pm$0.06 & 93.44$\pm$0.07 \\
& \textbf{GraphCodeBERT} & 92.45$\pm$0.09 & \cellcolor[HTML]{C0C0C0}95.23$\pm$0.11 & 93.17$\pm$0.13 \\\hline
\end{tabular}
}
\vspace{-3mm}
\end{table}

Then, we study how the hyperparameter $\alpha$ in Beta distribution for the determination of $\lambda$ in Eq.~(\ref{eqn:mixup}) influences the performance of \mixcode. The $\alpha$ can be set from 0 to $\infty$, but it is impractical to traverse all the possibilities. We follow the original Mixup setting where $\alpha = 0.2$ is the recommended setting and study $\alpha$ ranging from 0.05 to 0.5. Table \ref{tab:rq2_acc_alpha} shows both the test accuracy and robustness of trained models. Note that when setting the $\alpha$ as 0.5, the SeqofToken-based models are always poor (with less than 1\% accuracy). We conjecture that when $\alpha$ and $\lambda$ become bigger, the mixed code is meaningless, and the model cannot learn anything from the data. A deeper analysis of this interesting phenomenon will be our future work. From the remaining results, we can see that \mixcode can beat the standard training and basic data augmentation in most cases regardless of the setting of $\alpha$. The results demonstrate that a smaller $\alpha$ can produce a better model. In 13 (out of 18) cases, and 15 (out of 18) cases, $\alpha = 0.05$ or $\alpha = 0.1$ can produce models with higher accuracy and better robustness.


\begin{table*}[ht]
\centering
\caption{Results of \mixcode using Ori+Ref strategy for Mixup with different $\alpha$. The value with a gray background indicates the best result. The higher, the better.}
\label{tab:rq2_acc_alpha}
\centering
\resizebox{0.95\textwidth}{!}{
\begin{tabular}{llcccccc|cccccc}
\toprule
\textbf{Dataset} & \textbf{DNN} & $\alpha=$\textbf{0.05} & $\alpha=$\textbf{0.1} & $\alpha=$\textbf{0.2} & $\alpha=$\textbf{0.3} & $\alpha=$\textbf{0.4} & $\alpha=$\textbf{0.5} & $\alpha=$\textbf{0.05} & $\alpha=$\textbf{0.1} & $\alpha=$\textbf{0.2} & $\alpha=$\textbf{0.3} & $\alpha=$\textbf{0.4} & $\alpha=$\textbf{0.5} \\ \hline
\multicolumn{1}{c}{\textbf{}} &  & \multicolumn{6}{c|}{\textbf{Test Accuracy}} & \multicolumn{6}{c}{\textbf{Robustness}} \\ \hline
 & \textbf{BagofToken} & 82.08$\pm$0.09 & \cellcolor[HTML]{C0C0C0}82.32$\pm$0.07 & 82.19$\pm$0.05 & 81.76$\pm$0.05 & 81.15$\pm$0.03 & 80.92$\pm$0.05 & 77.63$\pm$0.06 & \cellcolor[HTML]{C0C0C0}78.17$\pm$0.05 & 78.15$\pm$0.03 & 77.94$\pm$0.04 & 77.90$\pm$0.04 & 77.28$\pm$0.03 \\
 & \textbf{SeqofToken} &  \cellcolor[HTML]{C0C0C0}95.23$\pm$0.28 & 94.52$\pm$0.02 & 93.20$\pm$0.26 & 90.84$\pm$0.65 & 90.55$\pm$0.61 & - & \cellcolor[HTML]{C0C0C0}86.66$\pm$0.25 & 85.60$\pm$0.01 & 82.96$\pm$0.03 & 81.43$\pm$0.01 & 81.18$\pm$0.02 & - \\
 & \textbf{CodeBERT} & 96.39$\pm$0.04 & 96.98$\pm$0.09 & \cellcolor[HTML]{C0C0C0}97.02$\pm$0.06 & 95.46$\pm$0.19 & 95.02$\pm$0.21 & 94.11$\pm$0.24 & \cellcolor[HTML]{C0C0C0}93.31$\pm$0.08 & 93.17$\pm$0.11 & 93.08$\pm$0.05 & 92.68$\pm$0.04 & 92.53$\pm$0.08 & 91.97$\pm$0.03 \\
\multirow{-4}{*}{\textbf{JAVA}} & \textbf{GraphCodeBERT} & 97.03$\pm$0.14 & \cellcolor[HTML]{C0C0C0}97.16$\pm$0.12 & 97.14$\pm$0.16 & 96.03$\pm$0.21 & 96.32$\pm$0.24 & 95.81$\pm$0.29 & \cellcolor[HTML]{C0C0C0}94.28$\pm$0.11 & 94.07$\pm$0.16 & 93.87$\pm$0.12 & 93.13$\pm$0.17 & 94.01$\pm$0.11 & 92.83$\pm$0.09 \\ \hline
 & \textbf{GGNN} & \cellcolor[HTML]{C0C0C0}65.44$\pm$0.29 & 65.42$\pm$0.35 & 65.31$\pm$0.26 & 65.19$\pm$0.33 & 64.78$\pm$0.46 & 64.65$\pm$0.25 & \cellcolor[HTML]{C0C0C0}55.61$\pm$0.03 & 55.01$\pm$0.01 & 54.88$\pm$0.04 & 54.32$\pm$0.03 & 54.08$\pm$0.02 & 53.97$\pm$0.03 \\
 & \textbf{GCN} & 64.12$\pm$0.31 & \cellcolor[HTML]{C0C0C0}64.18$\pm$0.44 & 64.03$\pm$0.34 & 63.89$\pm$0.28 & 63.76$\pm$0.38 & 63.56$\pm$0.24 & \cellcolor[HTML]{C0C0C0}54.81$\pm$0.01 & 54.03$\pm$0.03 & 53.97$\pm$0.04 & 53.42$\pm$0.05 & 53.08$\pm$0.02 & 52.78$\pm$0.03 \\
 & \textbf{GAT} & 64.03$\pm$0.28 & \cellcolor[HTML]{C0C0C0}64.09$\pm$0.36 & 63.78$\pm$0.27 & 63.58$\pm$0.31 & 63.23$\pm$0.29 & 63.07$\pm$0.22 & \cellcolor[HTML]{C0C0C0}53.92$\pm$0.01 & 53.81$\pm$0.03 & 53.26$\pm$0.02 & 52.97$\pm$0.01 & 52.69$\pm$0.02 & 52.35$\pm$0.04 \\
 & \textbf{CodeBERT} & \cellcolor[HTML]{C0C0C0}72.98$\pm$0.03 & 72.96$\pm$0.05 & 72.51$\pm$0.16 & 71.84$\pm$0.25 & 71.37$\pm$0.15 & 70.26$\pm$0.27 & 66.12$\pm$0.04 & \cellcolor[HTML]{C0C0C0}66.15$\pm$0.09 & 65.87$\pm$0.06 & 65.41$\pm$0.04 & 65.87$\pm$0.08 & 65.03$\pm$0.08 \\
\multirow{-5}{*}{\textbf{CodReq1}} & \textbf{GraphCodeBERT} & 73.11$\pm$0.14 & \cellcolor[HTML]{C0C0C0}73.34$\pm$0.13 & 72.67$\pm$0.19 & 72.19$\pm$0.22 & 71.62$\pm$0.25 & 70.34$\pm$0.29 & \cellcolor[HTML]{C0C0C0}66.23$\pm$0.11 & 65.91$\pm$0.13 & 65.36$\pm$0.18 & 64.78$\pm$0.21 & 64.85$\pm$0.17 & 64.21$\pm$0.21 \\ \hline
 & \textbf{BagofToken} & 67.14$\pm$0.09 & 67.88$\pm$0.07 & 68.22$\pm$0.06 & 68.61$\pm$0.05 & 68.53$\pm$0.08 & \cellcolor[HTML]{C0C0C0}68.72$\pm$0.06 & 63.31$\pm$0.05 & 63.65$\pm$0.02 & 64.58$\pm$0.01 & 64.74$\pm$0.03 & \cellcolor[HTML]{C0C0C0}64.82$\pm$0.02 & 64.73$\pm$0.01 \\
 & \textbf{SeqofToken} & 84.47$\pm$0.26 & \cellcolor[HTML]{C0C0C0}85.00$\pm$0.20 & 84.31$\pm$0.05 & 83.50$\pm$0.05 & 82.89$\pm$0.34 & - &  \cellcolor[HTML]{C0C0C0}83.53$\pm$0.16 & 82.94$\pm$0.02 & 82.17$\pm$0.01 & 81.24$\pm$0.02 & 80.81$\pm$0.03 & - \\
 & \textbf{CodeBERT} & \cellcolor[HTML]{C0C0C0}96.82$\pm$0.11 & 96.79$\pm$0.06 & 96.04$\pm$0.11 & 95.22$\pm$0.18 & 94.87$\pm$0.31 & 94.08$\pm$0.23 & 92.11$\pm$0.07 & \cellcolor[HTML]{C0C0C0}92.16$\pm$0.08 & 92.09$\pm$0.09 & 91.87$\pm$0.07 & 91.44$\pm$0.09 & 91.01$\pm$0.03 \\
\multirow{-4}{*}{\textbf{Python800}} & \textbf{GraphCodeBERT} & 97.10$\pm$0.06 & 97.09$\pm$0.08 & 97.01$\pm$0.14 & \cellcolor[HTML]{C0C0C0}97.13$\pm$0.09 & 96.27$\pm$0.24 & 95.67$\pm$0.13 & 94.87$\pm$0.03 & 94.56$\pm$0.07 & \cellcolor[HTML]{C0C0C0}94.95$\pm$0.11 & 94.02$\pm$0.13 & 93.24$\pm$0.19 & 92.67$\pm$0.13 \\ \hline
 & \textbf{GGNN} & 88.01$\pm$0.21 & \cellcolor[HTML]{C0C0C0}88.22$\pm$0.18 & 86.52$\pm$0.19 & 86.89$\pm$0.23 & 87.85$\pm$0.35 & 87.35$\pm$0.32 & 72.49$\pm$0.04 & \cellcolor[HTML]{C0C0C0}72.81$\pm$0.03 & 71.07$\pm$0.02 & 71.16$\pm$0.02 & 71.89$\pm$0.01 & 71.60$\pm$0.04 \\
 & \textbf{GCN} & 84.10$\pm$0.16 & \cellcolor[HTML]{C0C0C0}84.12$\pm$0.17 & 84.08$\pm$0.26 & 84.06$\pm$0.33 & 83.63$\pm$0.25 & 82.98$\pm$0.29 & \cellcolor[HTML]{C0C0C0}67.91$\pm$0.01 & 67.73$\pm$0.03 & 67.38$\pm$0.02 & 67.32$\pm$0.04 & 66.93$\pm$0.03 & 66.04$\pm$0.01 \\
 & \textbf{GAT} & \cellcolor[HTML]{C0C0C0}82.76$\pm$0.22 & 81.30$\pm$0.12 & 80.95$\pm$0.25 & 81.28$\pm$0.37 & 81.12$\pm$0.28 & 81.18$\pm$0.25 & \cellcolor[HTML]{C0C0C0}67.17$\pm$0.02 & 66.49$\pm$0.01 & 65.61$\pm$0.03 & 66.39$\pm$0.02 & 66.02$\pm$0.01 & 66.04$\pm$0.03 \\
 & \textbf{CodeBERT} & 95.78$\pm$0.12 & 97.57$\pm$0.37 & 96.69$\pm$0.19 & 95.94$\pm$0.22 & \cellcolor[HTML]{C0C0C0}97.85$\pm$0.24 & 95.59$\pm$0.33 & \cellcolor[HTML]{C0C0C0}95.52$\pm$0.06 & 95.41$\pm$0.06 & 95.01$\pm$0.05 & 95.37$\pm$0.03 & 94.86$\pm$0.03 & 94.16$\pm$0.04 \\
\multirow{-5}{*}{\textbf{Refactory}} & \textbf{GraphCodeBERT} & 98.03$\pm$0.24 & 98.16$\pm$0.21 & \cellcolor[HTML]{C0C0C0}98.47$\pm$0.25 & 98.23$\pm$0.27 & 97.92$\pm$0.21 & 97.44$\pm$0.23 & 95.31$\pm$0.13 & 95.23$\pm$0.11 & 95.17$\pm$0.17 & 94.21$\pm$0.27 & \cellcolor[HTML]{C0C0C0}95.53$\pm$0.31 & 94.07$\pm$0.29 \\ \bottomrule
\end{tabular}
}
\vspace{-3mm}
\end{table*}


\noindent\colorbox{gray!20}{\framebox{\parbox{0.96\linewidth}{
\textbf{Answer to RQ2}: Compared to single-type (only original or transformed) code mixing, using both types (original and transformed) code is the best strategy for \mixcode to train more accurate and robust models. A small $\alpha$ (e.g., 0.05 and 0.1) value is recommended for \mixcode.}}}

\subsection{RQ3: Impact of Refactoring Methods}

The refactoring method that determines the quality of transformed code is an important component in \mixcode. In RQ1 and RQ2, we consider randomly selecting refactoring methods from all the possibilities to prepare the transformed data. However, it is still unclear how these refactor methods influence the performance of \mixcode. Therefore, in this research question, we train the model by using each refactoring method separately under the best Mixup strategy \textit{Ori+Ref} proved in RQ2 to rank the refactoring methods based on the performance of the trained model. Then, we evenly split the 18 refactoring methods based on the ranking into good (high ranking) and poor (low ranking), two sets. We only consider two types of combinations because it is hard to consider all the situations. Afterward, we perform \mixcode again using these two sets, respectively. In this manner, we try to explore if there is a chance to further improve the \mixcode by using a better refactoring method combination. Here, we choose two models for our study, BagofToken for the problem classification task and GGNN for the bug detection task. \mixcode makes the best improvement in these two models, it is easier to amplify the difference.

\begin{table}[]
\centering
\caption{Refactoring methods selection (test accuracy). Good/poor: \mixcode using the best/worst 9 refactoring methods, respectively. Baseline: \mixcode using all 18 methods. The best method of 18 for each Dataset is highlighted with a gray background.}
\centering
\resizebox{\columnwidth}{!}{
\begin{tabular}{lcccc}
\toprule
\multirow{2}{*}{\textbf{Refactoring Method}} & \textbf{JAVA250} & \textbf{CodRep1} & \textbf{Python800} & \textbf{Refactory}\\ \cline{2-5}
 & \textbf{BagofToken} & \textbf{GGNN} & \textbf{BagofToken} & \textbf{GGNN} \\ \hline
\textbf{API Renaming} & 86.91$\pm$0.04 & 65.54$\pm$0.22 & 68.23$\pm$0.08 & 88.43$\pm$0.22 \\
\textbf{Arguments Adding} & \cellcolor[HTML]{C0C0C0}87.24$\pm$0.04 & 65.72$\pm$0.25 & 68.15$\pm$0.07 & 88.81$\pm$0.22 \\
\textbf{Argument Renaming} & 86.74$\pm$0.06 & 65.62$\pm$0.25 & 68.08$\pm$0.07 & 88.52$\pm$0.32 \\
\textbf{Dead For Adding} & 82.61$\pm$0.04 & 65.70$\pm$0.31 & 67.23$\pm$0.06 & 89.35$\pm$0.44 \\
\textbf{Dead If Adding} & 83.91$\pm$0.05 & 65.82$\pm$0.32 & 67.58$\pm$0.09 & 89.34$\pm$0.33 \\
\textbf{Dead If Else Adding} & 83.42$\pm$0.06 & 65.76$\pm$0.34 & 67.24$\pm$0.07 & 90.02$\pm$0.23 \\
\textbf{Dead Switch Adding} & 83.46$\pm$0.03 & 65.75$\pm$0.22 & - & - \\
\textbf{Dead While Adding} & 83.91$\pm$0.05 & 65.77$\pm$0.21 & 67.57$\pm$0.07 & 89.05$\pm$0.34 \\
\textbf{Duplication} & 85.61$\pm$0.06 & 65.44$\pm$0.33 & 67.32$\pm$0.06 & 88.36$\pm$0.21 \\
\textbf{Filed Enhancement} & 86.55$\pm$0.06 & 66.06$\pm$0.32 & 68.26$\pm$0.06 & 89.54$\pm$0.32 \\
\textbf{For Loop Enhancement} & 86.71$\pm$0.07 & 65.98$\pm$0.25 & 68.47$\pm$0.07 & 89.89$\pm$0.33 \\
\textbf{If Enhancement} & 86.35$\pm$0.03 & \cellcolor[HTML]{C0C0C0}66.35$\pm$0.23 & 68.26$\pm$0.05 & \cellcolor[HTML]{C0C0C0}91.04$\pm$0.33 \\
\textbf{Local Variable Adding} & 85.70$\pm$0.04 & 65.73$\pm$0.21 & 67.17$\pm$0.09 & 88.33$\pm$0.23 \\
\textbf{Local Variable Renaming} & 85.82$\pm$0.04 & 65.60$\pm$0.21 & 67.26$\pm$0.09 & 88.34$\pm$0.24 \\
\textbf{Method Name Renaming} & 87.02$\pm$0.05 & 65.04$\pm$0.43 & \cellcolor[HTML]{C0C0C0}68.59$\pm$0.06 & 88.32$\pm$0.22 \\
\textbf{Plus Zero} & 86.93$\pm$0.06 & 65.83$\pm$0.43 & 67.96$\pm$0.06 & 89.06$\pm$0.33 \\
\textbf{Print Adding} & 86.36$\pm$0.03 & 66.01$\pm$0.21 & 67.89$\pm$0.07 & 88.86$\pm$0.24 \\
\textbf{Return Optimal} & 85.85$\pm$0.06 & 65.53$\pm$0.34 & 67.28$\pm$0.08 & 88.33$\pm$0.22 \\ \hline
\textbf{Good (9)} & 86.34$\pm$0.06 & 65.72$\pm$0.24 & 68.03$\pm$0.06 & 89.97$\pm$0.17 \\ 
\textbf{Poor (9)} & 82.63$\pm$0.05 & 63.32$\pm$0.19 & 67.92$\pm$0.08 & 88.29$\pm$0.11 \\ 
\textbf{Baseline (18)} & 82.32$\pm$0.07 & 65.42$\pm$0.35 & 68.27$\pm$0.08 & 88.22$\pm$0.18 \\ \bottomrule
\end{tabular}
}
\label{tab:rq3_refactoring_acc}
\vspace{-5mm}
\end{table}

Table \ref{tab:rq3_refactoring_acc} presents the test accuracy of trained models using different individual program refactoring methods on the original test data. Surprisingly, \mixcode using a single refactoring method can produce more accurate models than using all the refactoring methods, and the gap can be up to 4.92\% (comparing \textit{Arguments Adding} to \textit{Baseline} in Java language). Then, the refactoring method with accuracy greater than 86.00\%, 65.75\%, 68.00\%, and 89.00\% from JAVA250, CodRep1, Python800, and Refactory is selected in the \textit{Good}, the others are put in the \textit{Poor}. We observe that considering only the test accuracy, compared to using \textit{Poor}, using \mixcode with refactoring methods from \textit{Good} can train a more accurate model. On average, \textit{Good} outperforms \textit{Poor} with 1.98\% test accuracy improvement. These results reveal that using a smartly selected single refactoring method is enough for \mixcode if we only care about the test accuracy of the trained model. Additionally, combining better-refactoring methods (which have higher single-test accuracy) can build a more effective \mixcode.

\begin{table}[]
\caption{Refactoring methods selection (robustness). Good/poor: \mixcode using the best/worst 9 refactoring methods, respectively. Baseline: \mixcode using all 18 methods. The best method of 18 for each Dataset is highlighted with a gray background.}
\centering
\resizebox{\columnwidth}{!}{
\begin{tabular}{lcccc}
\toprule
\multirow{2}{*}{\textbf{Refactoring Method}} & \textbf{JAVA250} & \textbf{CodRep1} & \textbf{Python800} & \textbf{Refactory}\\ \cline{2-5}
 & \textbf{BagofToken} & \textbf{GGNN} & \textbf{BagofToken} & \textbf{GGNN} \\ \hline
\textbf{API Renaming} & 43.61$\pm$0.02 & 55.06$\pm$0.02 & 40.11$\pm$0.02 & 66.65$\pm$0.02 \\
\textbf{Arguments Adding} & 42.72$\pm$0.03 & 55.09$\pm$0.01 & 39.05$\pm$0.01 & 66.68$\pm$0.02 \\
\textbf{Argument Renaming} & 43.27$\pm$0.04 & 55.04$\pm$0.01 & 40.08 $\pm$0.01 & 66.58$\pm$0.01 \\
\textbf{Dead For Adding} & 45.45$\pm$0.02 & 55.15$\pm$0.02 & 45.43$\pm$0.04 & 68.10$\pm$0.01 \\
\textbf{Dead If Adding} & \cellcolor[HTML]{C0C0C0}68.34$\pm$0.03 & 55.16$\pm$0.03 & \cellcolor[HTML]{C0C0C0}56.74$\pm$0.01 & 68.20$\pm$0.03 \\
\textbf{Dead If Else Adding} & 68.27$\pm$0.01 & \cellcolor[HTML]{C0C0C0}55.19$\pm$0.02 & 56.34$\pm$0.02 & \cellcolor[HTML]{C0C0C0}68.34$\pm$0.02 \\
\textbf{Dead Switch Adding} & 48.71$\pm$0.03 & 55.11$\pm$0.01 & - & - \\
\textbf{Dead While Adding} & 65.33$\pm$0.02 & 55.17$\pm$0.03 & 55.44$\pm$0.02 & 68.09$\pm$0.02 \\
\textbf{Duplication} & 43.29$\pm$0.03 & 55.03$\pm$0.03 & 40.88$\pm$0.02 & 66.67$\pm$0.01 \\
\textbf{Filed Enhancement} & 44.11$\pm$0.01 & 55.10$\pm$0.04 & 42.14 $\pm$0.02 & 66.59$\pm$0.02 \\
\textbf{For Loop Enhancement} & 42.42$\pm$0.03 & 55.09$\pm$0.02 & 41.29$\pm$0.02 & 66.64$\pm$0.02 \\
\textbf{If Enhancement} & 42.46$\pm$0.04 & 55.05$\pm$0.02 & 41.03$\pm$0.02 & 66.39$\pm$0.03 \\
\textbf{Local Variable Adding} & 43.34$\pm$0.02 & 55.08$\pm$0.02 & 40.78$\pm$0.03 & 66.32$\pm$0.02 \\
\textbf{Local Variable Renaming} & 44.44$\pm$0.04 & 55.09$\pm$0.03 & 40.67$\pm$0.02 & 66.34$\pm$0.01 \\
\textbf{Method Name Renaming} & 44.31$\pm$0.01 & 55.06$\pm$0.02 & 41.13$\pm$0.02 & 66.64$\pm$0.02 \\
\textbf{Plus Zero} & 43.16$\pm$0.01 & 55.06$\pm$0.02 & 40.65$\pm$0.02 & 66.43$\pm$0.02 \\
\textbf{Print Adding} & 44.27$\pm$0.02 & 55.03$\pm$0.04 & 41.08$\pm$0.03 & 66.67$\pm$0.02 \\
\textbf{Return Optimal} & 42.63$\pm$0.03 & 55.08$\pm$0.02 & 39.87$\pm$0.02 & 66.47$\pm$0.01 \\ \hline
\textbf{Good (9)} & 43.50$\pm$0.03 & 55.21$\pm$0.03 & 56.47$\pm$0.02 & 72.91$\pm$0.02 \\ 
\textbf{Poor (9)} & 78.19$\pm$0.02 & 52.09$\pm$0.02 & 63.55$\pm$0.01 & 71.47$\pm$0.01 \\
\textbf{Baseline (18)} & 78.17$\pm$0.05 & 55.01$\pm$0.01 & 63.65$\pm$0.02 & 72.81$\pm$0.03 \\ \bottomrule
\end{tabular}
}
\label{tab:rq3_refactoring_robust}
\vspace{-5.5mm}
\end{table}

Moving to the robustness, table \ref{tab:rq3_refactoring_robust} presents the results of trained models. First, compared with the standard training (in table \ref{tab:rq1_acc}), \mixcode with a single refactoring method can still produce more robust models. Only one case, \textit{Local Variable Adding - Refactory}, has lower robustness than standard training. Then, we compare single-method and multiple-method combinations. Different from the pure test accuracy, the results show that \mixcode with a single refactoring method can not outperform using multiple methods, and the difference gap is huge. For example, in JAVA250, the 
robustness of using a single method ranges from 42.42\% to 68.34\%, but the robustness of using \textit{Poor} can be 78.19\%, where around 10\% robustness difference appeared.  It is reasonable since if we force the model to learn one specific code refactoring, the generalization of the trained model could be quite low. Then, we compare \textit{Good} and \textit{Poor}. Surprisingly, in half of the cases (2 out of 4), \textit{Poor} has higher robustness than \textit{Good}. It is not the case that models trained by using better-refactoring methods (with higher original test accuracy) always have better robustness. This finding reveals a trade-off between original test accuracy and robustness when considering the refactoring methods for \mixcode.


\noindent\colorbox{gray!20}{\framebox{\parbox{0.96\linewidth}{
\textbf{Answer to RQ3}: Using a specific refactoring method (depending on the dataset and DNN), \mixcode can produce models with high test accuracy but has to scarify the robustness. Utilizing multiple methods to enrich the diversity in data remains the best solution to adjust the trade-off between test accuracy and robustness.}}}

\section{Discussion}
\label{sec:discussion}
In this section, we discuss the limitation of our work, potential future research directions, and the threats to validity. 

\subsection{Limitation \& Future Directions}
\noindent
\textbf{Limitation.} In Section \ref{seq:rq2_2}, we observed that the setting of hyperparameter $\alpha$ could affect the performance of \mixcode. And in the worst case, e.g., $\alpha=0.5$, some models can learn nothing from the mixed data. The potential reason is still unclear, raising concerns about using \mixcode. However, it is possible to bypass this limitation by using \mixcode with a smaller $\alpha$.

\noindent
\textbf{Future directions.} Existing works for source code analysis mainly focus on proposing new code representation and code embedding approaches~\cite{ma2021graphcode2vec,guo2020graphcodebert} or studying how to utilize large pre-trained language models for downstream code tasks~\cite{mastropaolo2021studying}. Only limited works consider the importance of the quality of training data~\cite{sun2022importance} or the importance of training strategies. There are still many opportunities in source code analysis for better program coding. We highlight 
two potential research directions for future exploration.

\textbf{1)} In terms of \mixcode, we investigate the impact of different refactoring methods and combinations on both the test accuracy and robustness of FNN, CNN, GGNN, GCN, GAT, CodeBERT, and GraphCodeBERT. Despite that \mixcode brings better performance, there is still room to improve. The straightforward research direction is to design an adaptive method to find the best combination of refactoring methods.

\textbf{2)} In terms of Mixup, a series of works propose different variants of the original Mixup. Therefore, in addition to using raw code vectors to do code mixing, other options, such as the raw source code and the embedding vectors, can also be used as the input of the Mixup approach.


\subsection{Threats to Validity}
The internal threat to validity comes from the implementation of standard training, basic data augmentation, and \mixcode. The training for the classification tasks (JAVA250, Python800) is taken from the project CodeNet~\cite{puri2021codenet}, and the defect detection tasks (CodRep1, Refactory) adopt from~\cite{hu2019re, chen2018codrep, zhong2015empirical}. The implementation of Mixup comes from its original release~\cite{zhang2017mixup}. The 18 refactoring methods for the Java language are from~\cite{pour2021search, wei2021cocofuzzing}, and we adapt the implementation to the Python language. 

The external threats to validity lie in the selected source code tasks, datasets, DNNs, and refactoring methods. We consider two different tasks (problem classification and bug detection) in the study and include two datasets for each task. Particularly, we include two popular programming languages (Java and Python) for software developers. Remarkably, we utilize seven types of deep neural networks, including the most famous pre-trained programming language models. For the refactoring methods, we cover the most common ones from the literature.

The construct threats to validity mainly come from the parameters of \mixcode, randomness, and evaluation measures. \mixcode only contains the parameter $\lambda$ that controls the weight of mixing two input instances. We follow the recommendation of the original Mixup algorithm and investigate the impact of this parameter. We observe that regardless of the parameter setting, \mixcode still outperforms the standard training and basic data augmentation. To reduce the impact of randomness, we repeat each experiment five times and report the average and standard deviation results. Finally, for evaluation measures, we consider both the accuracy of the original test data and the robustness of transformed test data. The latter one is specific for evaluating the generalization ability of DNNs.

\section{Related work}
\label{sec:relatedworks}
We consider related works from three aspects, source code learning enhancement, data augmentation for source code analysis, and Mixup for image and text classification.

\subsection{Source Code Learning Enhancement}

The goal of our work is to improve the performance of source code-related models. Existing works targeting the same goal try to challenge this problem in different ways. 

\noindent
\textbf{Self-supervised learning.} Although the code model has significantly succeeded in software engineering applications, there are still some limitations. For example, some code models are built for solving a particular problem, thus, the related code representation is difficult to extend to other problems. To enhance the flexibility of code models, Bui et al.~\cite{bui2021infercode} proposed to utilize a self-supervised learning mechanism for code representation preparation. Specifically, it utilizes the identified sub-trees prediction as the self-supervised task to train the code representation. As a result, the code representation can be used in different downstream tasks.

\noindent
\textbf{Pre-trained models fine-tuning.} 
Reusing pre-trained models to solve code analysis problems is another straightforward method. Multiple models have been proposed. Kanade et al.~\cite{kanade2020learning} introduced CuBERT, the same model architecture as BERT\cite{devlin2018bert}, trained on Python source code. Buratti et al.~\cite{buratti2020exploring} used a transformer-based model that was trained on C language to build the pre-trained model C-BERT. Here, CuBERT and C-BERT were both trained by using a single programming language. More recently, some multi-programming language pre-trained models were proposed. Lu et al.~\cite{lu2021codexglue} provided CodeGPT, which was trained on Python and Java corpora. Feng et al.~\cite{feng2020codebert} presented CodeBERT, a pre-trained model that learns information from six programming languages and natural language text. To further capture the semantic structure of code, Guo et al.~\cite{guo2020graphcodebert} proposed GraphCodeBERT, which uses data flow information of code in the pre-training stage. Different from these works, \mixcode mainly focuses on improving the model training process from the perspective of data augmentation. As a result, \mixcode can be generalized to any code classification task regardless of pre-trained models.

\subsection{Data Augmentation for Source Code Analysis} 

Data augmentation has achieved tremendous success in the CV and NLP fields~\cite{shorten2019survey,feng2021survey}. Recently, due to the similar processing workflow of NLP data and source code, researchers devoted considerable effort to applying this technique to improve the performance of the code model. The widely studied data augmentation method, adversarial training~\cite{goodfellow2014explaining}, has been studied in code learning. Simply, adversarial training generates a set of adversarial examples and combines them with the original training data to increase the data volume. For instance, Zhang et al.~\cite{zhang2020training} generated adversarial examples based on the Metropolis-Hastings modifier (MHM) algorithm~\cite{zhang2020generating}. Mi et al.~\cite{mi2021effectiveness} generated the synthetic data from Auxiliary Classifier generative adversarial networks (GANs) to increase the data size. Unlike the above works, we generate new data by refactoring the programs and then apply Mixup to enrich the volume and diversity of the training dataset.

\subsection{Mixup for CV and NLP }

Compared to the above-mentioned data augmentation methods that increase the data volume by combining adversarial examples and the original training data, Mixup~\cite{zhang2017mixup} linearly mixes existing training data to increase the diversity of learned information by the model. Recently, multiple variants of Mixup have been proposed~\cite{guo2019augmenting,sun2020mixup,chen2020mixtext,zhang2020seqmix,walawalkar2020attentive,uddin2020saliencymix,qin2020resizemix,kim2020puzzle,kim2021co,verma2019manifold}. Yun et al.~\cite{yun2019cutmix} applied Dropout~\cite{srivastava2014dropout} into Mixup, and proposed a mixing strategy based on the patch of the image. Besides, Liu et al.~\cite{liu2021unveiling} introduced the Automatic Mixup (AutoMix) strategy to balance the mixing policies and optimization complexity. Although the original Mixup is proposed for image data, researchers have extended the Mixup to support other types of data. For text data, Yoon et al.~\cite{yoon2021ssmix} synthesized the new text data from two raw input data by replacing the hidden vectors based on span-based mixing. Wang et al.~\cite{wang2021mixup} provided a two-stage Mixup framework for graph data. One is mixing the feature of node neighbors, and another is mixing the feature of the entire graph. Like Mixup for text and graph data classification, we augment the input code data in the vector space for source code classification. However, the difference is that we consider the code with sequence and graph representation, and we mix the data with their transformed version.

\section{Conclusion}
\label{sec:conclusion}

This paper presented \mixcode, the first data augmentation framework for source code analysis, to enhance model performance without collecting or labeling new code. Specifically, \mixcode supports 18 types of refactoring methods (extensible with new ones) that generate transformed code. To evaluate the effectiveness of \mixcode, we conducted extensive experiments on two important code tasks (problem classification and bug detection) and seven DNN architectures. Experimental results demonstrated \mixcode outperforms the basic data augmentation baseline by up to 6.24\% accuracy and 26.06\% robustness improvement. The configuration study proved that linearly mixing the original and transformed code achieves the best performance of \mixcode. 
\section{Acknowledgement}
This research is supported in part by JSPS KAKENHI Grant No. JP19H04086, Japan. Qiang Hu is also supported by the Luxembourg National Research Funds (FNR) through CORE project C18/IS/12669767/STELLAR/LeTraon.

\bibliographystyle{IEEEtran}
\bibliography{reference}

\end{document}